\documentclass[journal]{IEEEtran}
\usepackage{fancyhdr} 
\usepackage{lastpage} 
\usepackage{extramarks} 
\usepackage{graphicx} 
\usepackage{lipsum} 
\usepackage{amssymb}
\usepackage{color}
\usepackage{amsmath}
\usepackage[final]{pdfpages}
\usepackage{paralist}
\usepackage[usenames,dvipsnames]{xcolor}
\usepackage{amsthm,amsmath,amsfonts,ed,empheq}
\usepackage{xspace}
\usepackage{url}
\usepackage{multirow}
\usepackage{footnote}
\usepackage{amssymb,amsmath,mathtools,fixmath}
\usepackage{algorithmicx,algorithm}
\usepackage{algpseudocode} 
\usepackage{cases}
\usepackage{cite}
\usepackage{graphicx,amsfonts,psfrag,url,booktabs}
\usepackage{comment}
\usepackage{cite}
\usepackage{xcolor}
\usepackage[caption=false,font=footnotesize]{subfig}
\usepackage{makecell}
\usepackage{psfrag}
\usepackage{tabularx}
\usepackage[table]{colortbl}
\usepackage{booktabs}
\usepackage{balance}
\usepackage{algorithm}
\usepackage{algpseudocode}
\usepackage{hyperref}
\usepackage{flushend}

\newcommand{\A}{\mathbold{A}}
\newcommand{\B}{\mathbold{B}}
\newcommand{\C}{\mathbold{C}}
\newcommand{\D}{\mathbold{D}}
\newcommand{\F}{\mathbold{F}}

\newcommand{\K}{\mathbold{K}}
\newcommand{\I}{\mathbold{I}}
\newcommand{\J}{\mathbold{J}}

\renewcommand{\L}{\mathbold{L}}
\renewcommand{\S}{\mathbold{S}}

\renewcommand{\P}{\mathbold{P}}
\renewcommand{\O}{\mathbold{O}}

\newcommand{\U}{\mathbold{U}}

\newcommand{\W}{\mathbold{W}}
\newcommand{\X}{\mathbold{X}}

\renewcommand{\a}{\mathbold{a}}

\renewcommand{\c}{\mathbold{c}}

\renewcommand{\v}{\mathbold{v}}

\renewcommand{\r}{\mathbold{r}}
\newcommand{\x}{\mathbold{x}}
\newcommand{\y}{\mathbold{y}}
\newcommand{\w}{\mathbold{w}}
\newcommand{\z}{\mathbold{z}}
\renewcommand{\u}{\mathbold{u}}

\newcommand{\tA}{\tilde{\mathbold{A}}}
\newcommand{\tB}{\tilde{\mathbold{B}}}
\newcommand{\1}{\mathbf{1}}
\newcommand{\bLam}{\mathbold{\Lambda}}

\newcommand{\bC}{\bar{\mathbold{C}}}

\newcommand{\Fcal}{\mathcal{F}}
\newcommand{\Scal}{\mathcal{S}}
\newcommand{\Gr}{\mathcal{G}}
\newcommand{\Vx}{\mathcal{V}}

\newcommand{\Ed}{\mathcal{E}}

\newcommand{\herm}{{\mathsf{H}}}
\newcommand{\transp}{^{\mathsf{T}}}
\newcommand{\hx}{\hat{\mathbold{x}}}
\newcommand{\tx}{\tilde{\mathbold{x}}}
\newcommand{\tu}{\tilde{\mathbold{u}}}
\newcommand{\tw}{\tilde{\mathbold{w}}}
\newcommand{\tF}{\tilde{\mathbold{F}}}

\newcommand{\Exp}{\mathbb{E}}
\newcommand{\bSigma}{\mathbold{\Sigma}}

\DeclareMathOperator{\tr}{tr}
\newcommand{\bPsi}{\mathbold{\Psi}}

\newcommand{\Cxp}{\mathbb{C}}
\newcommand{\del}{\partial}
\renewcommand{\tr}{\text{Tr}}

\newcommand{\MSE}{\textnormal{MSE~}}
\newcommand{\LS}{\textnormal{LS~}}

\newcommand{\vsp}{\vskip -0mm}
\newcommand{\vspone}{\vskip -1mm}

\newtheorem{theorem}{Theorem}

\newtheorem{proposition}{Proposition}
\newtheorem{definition}{Definition}
\newtheorem{corollary}{Corollary}
\newtheorem{remark}{Remark}
\newtheorem{assumption}{Assumption}

\newcommand{\rev}{\textcolor{black}}

\renewcommand\baselinestretch{.97}

\begin{document}
\title{Observing and Tracking Bandlimited Graph Processes}


\author{\IEEEauthorblockN{Elvin Isufi$^{*,\dagger}$, ~\IEEEmembership{Student Member,~IEEE}, Paolo Banelli$^{\dagger}$,~\IEEEmembership{Member,~IEEE}, Paolo Di Lorenzo$^{\dagger}$,~\IEEEmembership{Member,~IEEE}, and Geert Leus$^{*}$~\IEEEmembership{Fellow,~IEEE}\vspace{-.3cm}}
\thanks{$^{*}$Faculty of EEMCS, Delft University of Technology, 2628 CD Delft, The Netherlands. $^{\dagger}$ Department of Engineering, University of Perugia, Via G. Duranti 93, 06125, Perugia, Italy. The work of P. Di Lorenzo was supportend by the "Fondazione Cassa di Risparimio di Perugia".
E-mails: \{e.isufi-1; g.j.t.leus\}@tudelft.nl, \{paolo.banelli; paolodilorenzo\}@unipg.it. }}

%



\IEEEtitleabstractindextext{%
\begin{abstract}
One of the most crucial challenges in graph signal processing is the sampling of bandlimited graph signals, i.e., signals that are sparse in a well-defined graph Fourier domain. So far, the prior art is mostly focused on (sub)sampling single snapshots of graph signals ignoring their evolution over time. However, time can bring forth new insights, since many real signals like sensor measurements, biological, financial, and network signals in general, have intrinsic correlations in both domains. 

In this work, {we fill this lacuna} by jointly considering the graph-time nature of graph signals, named \emph{graph processes} for two main tasks: \emph{i)} observability of graph processes; and \emph{ii)} tracking of graph processes via Kalman filtering; both from a (possibly time-varying) subset of nodes.
A detailed mathematical analysis ratifies the proposed methods and provides insights into the role played by the different actors, such as the graph topology, the process bandwidth, and the sampling strategy. Moreover, (sub)optimal sampling strategies that jointly exploit the nature of the graph structure and graph process are proposed.

Several numerical tests on both synthetic and real data validate our theoretical findings and illustrate the performance of the proposed methods in coping with time-varying graph signals.
\end{abstract}

\begin{IEEEkeywords}
Graph signal processing, sampling on graphs, time-varying graph signals, observability, graph processes, Kalman filtering.\vspone\vspone\vspone
\end{IEEEkeywords}}

\maketitle

\IEEEdisplaynontitleabstractindextext

%

\section{Introduction}
\label{sec.intro}

Graph signal processing (GSP) \cite{Shuman2013} is a promising tool for analyzing network signals, such as measurements in sensor networks, {fMRI} data in brain networks, or personal preferences in opinion networks. It distinguishes itself from other methods by providing a specific notion of graph frequency \cite{Shuman2013,Sandryhaila2013} which allows for a harmonic analysis of graph signals. In this context, several works extended classical concepts, like filtering \cite{Sandryhaila2013, segarra2017optimal, Isufi2017ARMA, isufi2017filtering}, wavelets \cite{hammond2011wavelets}, filter-banks \cite{Narang2012,tanaka2014} and sampling \cite{chen2015samp,pesenson2008sampling,marques2016sampling,tsitsvero2016signals} to signals on graph.

\rev{The sampling theory for graph signals mainly relies on the observation that the latter are often sparse (i.e., bandlimited) in the graph frequency domain. Practical examples of bandlimited graph signals include temperature measurements, where adjacent sensors measure similar values, fMRI data of brain networks \cite{hu2016msdGS}, ratings in recommendation systems \cite{gomez2016netflix}, protein networks \cite{yamanishi2004protein}, and networks that exhibit a clustering behavior such as opinion networks. This characteristic is exploited in a number of works including \cite{wang2015local,narang2013localized,chen2015samp,marques2016sampling,tsitsvero2016signals} to propose signal reconstruction strategies from a few measurements.}

All the above works propose sampling and reconstruction strategies only for a single snapshot of the graph signal ignoring its time-varying nature. Time-varying graph signals, named \emph{graph processes}, are encountered in consecutive sensor measurements, biological signal evolution prone to stimuli, financial networks, and information diffusion over networks. \rev{In this work, we extend the graph signal sampling theory to a graph-time framework for the tasks of observing and tracking a bandlimited graph process, i.e., a graph process that has a sparse graph frequency content over time.} Before detailing the paper contributions, in the sequel, we highlight the differences with earlier works that dealt with similar problems.\vspone\vspone\vspone\vspone

\subsection{Related works}

Several recent works have recognized the importance of extending GSP to the temporal dimension. The works \cite{sandryhaila2014big,grassi2017time} focus on harmonic analysis for time-varying graph signals, while  \cite{isufi2016ARMA2D,Isufi2016Dim2} on graph-time filters. Here, we continue this extension for observing and tracking a graph process from a few measurements.

Observability of network processes has been considered in \cite{xue2016minimum,pequito2015minimum} for sensor placement and in \cite{pequito2014optimal} for designing observable topologies. While these findings are of particular importance, these works do not consider sampling strategies for observing the network process. Differently, we exploit GSP tools and in particular the bandlimited prior to bring the graph sampling theory into the temporal dimension. This allows us to derive by theoretical guarantees when a bandlimited network process is observable from a few measurements and to propose effective graph-time sampling schemes.

The bandlimited assumption has also been exploited in \cite{PDL2016_TSIPN,PDL2016_DLMS,didistributed,di2017adaptive} for tracking slowly time-varying graph signals from a few nodes.
However, since the main goal in these works is to develop sampling strategies for adaptive signal reconstruction, signal tracking comes as a byproduct and, therefore, without theoretical guarantees.
Here, we tackle this challenge and we propose sampling strategies directly to track the process by means of Kalman filtering (KF), which is known to achieve the optimal performance.

\rev{The works in \cite{mei2015signal,LoukasPredict} exploited the graph structure to improve the prediction accuracy of autoregressive (AR) and autoregressive moving average (ARMA) models. However, the findings in these works need all the data to stand. As we show later, this is not the case for the proposed KF.}

\rev{The work in \cite{romero2016kernel} casts the tracking of a graph process from limited measurements as a regularized-based interpolation problem on a larger graph. Subsequently, KF-like methods are used to alleviate the computational burden. We identify three main differences with the proposed framework. First, no conditions on the minimum number of required samples are provided in \cite{romero2016kernel}. Second, the sampling is performed uniformly at random, which is well-known to be a suboptimal choice. Finally, since our framework is not a regularized interpolation problem, we avoid the task of designing the regularizer and its weight to track the process.}

Differently from the above, KF has also been used to track network dynamics \cite{soule2005traffic,cattivelli2010diffusion}. However, in these cases, the graph embeds the communication links between the sensors. To the best of our knowledge, this is the first attempt that conciliates process tracking with a designed sampling strategy and with the nature of the graph process itself, i.e., its bandlimitedness.\vspone\vspone\vspone

\subsection{Contribution and notation}

We can divide the paper contributions into two parts.

\emph{i) Observability of graph processes (Section~\ref{sec.observ}):} We extend the sampling theory for graph signals to the observability of graph processes. We derive \emph{necessary} and \emph{sufficient} conditions for observing a bandlimited graph process from a subset of nodes. We propose two approaches for observability: \emph{i-a) observability with deterministic sampling (Section~\ref{subsec.det_obs})}, i.e., when the selected subset of nodes is chosen deterministically; and \emph{i-b) observability with random sampling (Section~\ref{subsec.rand_obs})}, i.e., when a subset of nodes is sampled with a given probability. We perform a mean-square error (MSE) analysis of the state estimation performance to show the connection between the graph topology, the process bandwidth, and the sampling set. Finally, we propose sampling techniques based on sparse sensing to pick the minimum number of samples such that a target MSE estimation performance is guaranteed. 

\emph{ii) Kalman filtering for graph processes (Section V):} We propose KF to track bandlimited graph processes that follow a predefined model. We first consider KF for time-varying models (Section~\ref{subs.KF}) and then extend our derivations to steady-state KF (Section~\ref{subsec_ssKF}), i.e., when the model is time-invariant. We derive \emph{necessary} conditions on the minimum required number of nodes for tracking the graph process and conciliate these conditions with those derived for observability. The MSE analysis, given by the posterior (or steady-state) error covariance matrix highlights the role played by the graph topology, the graph process bandwidth, and the sampling set in the tracking (steady-state) performance. A sparse sampling strategy, which ensures a predefined MSE tracking cost, is then used to sample the graph. 

Several numerical tests on synthetic and real data corroborate the theoretical findings and illustrate the performance of the proposed methods.

\textbf{Notation.} Scalar, column vector, and matrix variables are respectively indicated by plain letters $a$ ($A$), bold lowercase letters $\a$ and bold uppercase letters $\A$. $A_{i,j}$ is the $(i,j)$th element of $\A$, $\I_N$ is the $N \times N$ identity matrix, and $\mathbf{1}_N$ ($\mathbf{0}_N$) is the $N \times 1$ vector of all ones (zeros). diag$(\cdot)$ denotes the diagonal operator, i.e., $\A = \text{diag}(\a)$ is a diagonal matrix with $\a$ on the main diagonal and $\a = \text{diag}(\A)$ stores the diagonal of $\A$ in $\a$. Similarly, ${\A} = \text{blkdiag}(\A_1, \ldots, \A_N)$ is a block diagonal matrix containing $\A_i$ as the $i$th diagonal block. The pseudoinverse of $\A$ is denoted as $\A^\dagger$, its trace as $\tr(\A)$, and the matrix spectral norm as $\|\A\|$. For two sets $\mathcal{R}$ and $\Scal$, $|\mathcal{R}|$ is the cardinality of $\mathcal{R}$, $\mathcal{R} \subset \Scal$ denotes the subset operation, $\mathcal{R} \cup \Scal$ the union of the two sets, and $\Scal^c = \mathcal{R}\backslash\Scal$ the complementary set of $\Scal$ w.r.t. $\mathcal{R}$. The vector $\mathbf{1}_\mathcal{R}$ is the set indicator vector, whose $r$th entry is equal to one if $r \in \mathcal{R}$ and zero otherwise. $\lceil \cdot \rceil$ indicates the ceiling operator.

{Section~\ref{sec.background} covers the background information, while Section~\ref{sec.probl} formulates the problem under consideration. Section~\ref{sec.simulations} contains the numerical evaluation and Section~\ref{sec.concl} the conclusions. The proofs are in appendix.}


\section{Background}
\label{sec.background}

This section recalls the basics of GSP and graph signal reconstruction.

\textbf{Basics of GSP.} Consider an undirected graph $\Gr = (\Vx,\Ed)$, where $\Vx$ indicates the set of $N$ nodes and $\Ed$ the edge set. The nodes' connectivity is captured by the weighted adjacency matrix $\W$, where $W_{n,m}>0$ is the edge weight connecting the tuple $(n,m)$. $W_{n,m} = 0$ means that nodes $n$ and $m$ are not connected. The graph Laplacian matrix is $\L_{\text{d}} = \text{diag}(\W\1_N) - \W$. To each node $n$ a signal is associated, named the graph signal, $x_n \in \Cxp$. For convenience, we collect the graph signals in the vector $\x = [x_1, \ldots, x_N]^{\transp}$.

Both $\L_{\text{d}}$ and $\W$ are candidates for the graph shift operator matrix $\S$, an $N \times N$ matrix which plays a central role in the frequency analysis of graph signals \cite{taubiny2000geometric,Shuman2013,Sandryhaila2013}. Due to its symmetry, $\S$ is eigendecomposed as $\S = \U\bLam\U^{\herm}$ with $\U$ the eigenvector matrix and $\bLam$ the diagonal matrix containing the eigenvalues. \cite{Shuman2013,Sandryhaila2013,taubiny2000geometric} and references therein have advocated that the eigendecomposition of $\S$ carries the notion of frequency in the graph setting. Specifically, the projection of $\x$ onto the eigenspace of $\S$, i.e., $\hat{\x} = \U^{\herm}\x$ is named the graph Fourier transform of $\x$ and its inverse is $\x = \U\hat{\x}$. In this context, the diagonal matrix $\bLam$ contains the spectral support for $\hx$, i.e., the graph frequencies.

\rev{\textbf{Signal reconstruction.} The reconstruction of a graph signal $\x$ from its sampled version is related to the joint localization properties of $\x$ in the vertex and graph frequency domain \cite{tsitsvero2016signals}. For $\Scal \subseteq \Vx$ being a subset of vertices and $\C_{\Scal} = \text{diag}(\mathbf{1}_{\Scal})$ the respective set projection matrix, $\x$ is said to be perfectly localized over $\Scal$ if $\C_{\Scal}\x = \x$. Similarly, given a subset of graph frequency indices $\Fcal \subseteq \{1, \ldots, N\}$, we define the matrix $\U_\Fcal \in \mathbb{C}^{N\times |\Fcal|}$ containing the columns of $\U$ relative to the set $\Fcal$ and the bandlimiting operator $\C_\Fcal = \U_\Fcal\U_\Fcal^\herm$. Then, $\x$ is said to be perfectly localized over $\Fcal$ (or $\Fcal-$bandlimited) if $\C_\Fcal\x = \x$.}

\rev{In \cite{tsitsvero2016signals}, it has been shown that any $\Fcal-$bandlimited graph signal $\x$ can be perfectly recovered from samples collected over the set $\Scal$ if and only if
\begin{equation}
\label{eq.rec_x}
\|\C_{\Scal^c}\U_\Fcal\| < 1,
\end{equation}
i.e., there are no $\Fcal-$bandlimited graph signals perfectly localized over the complementary vertex set $\Scal^c = \Vx\backslash\Scal$. Here, $\C_{\Scal^c} = \I_N - \C_{\Scal}$ is the projection matrix onto $\Scal^c$.}

\section{Problem Statement}
\label{sec.probl}

In this section, we first reformulate the graph process as a linear system on graphs. Then, we define the bandlimited graph process and reformulate the system on graphs as a sparse representation in the GFT domain.\vspone\vspone\vspone

\subsection{Systems on graphs}
\label{subsec.gsd_ssmg}

Consider the $N-$state discrete linear time-varying system
\begin{subequations}
\label{eq.state_space}
\begin{align}
\label{eq.kf_equations}
\x_t &= \A_{t-1}\x_{t-1} + \B_{t-1}\u_{t-1}\\
\y_{t} &= \C_{\Scal_t}(\x_t+ \v_t),
\label{eq.measurement}
\end{align}
\end{subequations}
where $\x_t$ is the state vector containing the graph signal at time $t$, $\u_t$ is the input signal, and $\A_{t}$ and $\B_{t}$ are the time-varying state-transition and input matrices, respectively. $\y_{t} \in \Cxp^{N}$ is the measurement vector and $\C_{\Scal_t} = \text{diag}(c_{t,1}, \ldots, c_{t,N})$ is the sampling matrix with $c_{t,n} = 1$ if the $n$th node belongs to the instantaneous sampling set ${\Scal_t}$ defined as $\Scal_t = \{n \in \{1, \ldots, N\} | c_{t,n} = 1 \}$. $\v_t$ is white zero-mean noise with covariance matrix $\bSigma_v = \sigma_v^2\I_N$. 

Model \eqref{eq.state_space} comprises the following network processes.

\textbf{Signal diffusion.} For $\x_0$ being the initial signal state on the graph, its instantaneous diffused \cite{Kondor2002} realization is expressed through the exponential matrix product
\begin{equation}
\label{eq.inst_diff}
\x_t = e^{-w\L_{\text{d}} t}\x_0 = e^{-w\L_{\text{d}}}e^{-w\L_{\text{d}}(t-1)}\x_0 \triangleq \A\x_{t-1}
\end{equation}
where $w > 0$ is the diffusion rate and $\A = e^{-w\L_{\text{d}}}$ is the time-invariant state-transition matrix. In \eqref{eq.inst_diff} we can also incorporate an input $\u_{t-1}$ which may represent additional sources that become available at $t-1 > 0$. \rev{The diffusion model has found several practical applications including temperature diffusion, chemical substances dispersion, opinion propagation over networks \cite{Dittmer2001}, and brain signal analysis \cite{Tarun2018}.}

\rev{\textbf{Wave propagation.} The discretised wave equation on graphs \cite{friedman2004wave} follows the two-step recursion
\begin{equation}
\label{eq.wave_eq}
\w_t = (2\I_N - c^2\L_{\text{d}})\w_{t-1} - \w_{t-2},
\end{equation}
with initial state $\w_0$ and wave speed $c$. Recursion \eqref{eq.wave_eq} can be reformulated as~\eqref{eq.kf_equations} by defining $\x_t = [\w_{t-1}, \w_{t}]{\transp}$ with $\A = \big[[\I_N, (2\I_N-c^2\L_{\text{d}})]{\transp}, [\mathbf{0}_N{\mathbf{0}_N}{\transp}, \I_N]{\transp}\big]$ and $\w_{-1} = {\mathbf{0}_N}$. The latter is of practical interest for instance in seismic data \cite{grassi2017time}.}

\rev{\textbf{ARMA graph processes.} We denote an ARMA graph process as
\begin{equation}\label{eq.ARMAgp}
\x_t = f(\S)\x_{t-1} + g(\S)\u_{t-1},
\end{equation}
where $f(\S)$ and $g(\S)$ are matrix functions of $\S$ that share the eigenvectors with $\S$, such as polynomials of a given power \cite{mei2015signal,isufi2017autoregressivest,LoukasPredict}.}

\rev{A particular form of \eqref{eq.ARMAgp} is the first-order recursion
\begin{equation}
\label{eq.ss_arma1}
\x_t = -w\S\x_{t-1} + \u_0~\text{with}\quad \x_0 = \mathbf{0}_N,
\end{equation}
which for $0 < w < 1/\lambda_{\text{max}}(\S)$ reaches the steady-state
\begin{equation}\label{eq.ss_solarma1}
\x = \lim_{t \to \infty}\x_t = (\I_N + w\S)^{-1}\u_0.
\end{equation}
Expression \eqref{eq.ss_solarma1} is the solution of the so-called aggregate diffusion model and is exploited in image smoothing \cite{Zhang2008}, Tikhonov denoising \cite{Shuman2013}, and recommendation systems \cite{Ma2016}.} 

\subsection{Bandlimited systems on graphs}
\label{subsec.blimssmg}

To proceed with the graph Fourier analysis of \eqref{eq.state_space}, we define the following.

\begin{definition}
\label{def.blimGP} A graph process $\x_t$ with instantaneous GFT $\hx_t \!=\! \U^\herm\x_t$ is $\Fcal-$bandlimited if $\hx_t$ has non-zero frequency content only on a subset of graph frequency indices $\Fcal$.
\end{definition}
The set $\Fcal = \{n \in \{1, \ldots, N	\} | \hat{x}_{t,n} \neq 0, t \ge 0\}$ is considered to be a common \emph{time-invariant} set for all realizations of $\x_t$. Said differently, $\Fcal$ is the union of all instantaneous sets $\Fcal_t = \{n \in \{1, \ldots, N	\} | \hat{x}_{t,n} \neq 0\}$. We then write
\begin{equation}
\label{eq.sprse_freq}
\x_t = \U_\Fcal\tilde{\x}_t,
\end{equation}
where $\tilde{\x}_t \in \Cxp^{|\Fcal|}$ is the vector containing the entries of $\hx_t$ indicated by $\Fcal$.

We further assume the following. \vspone\vspone\vspone\vspone

\rev{\begin{assumption}\label{as.shift} The system evolution matrices $\A_t$ and $\B_t$ share the eigenvectors with the graph shift operator $\S$.
\end{assumption}}\vspone\vspone\vspone
\rev{\begin{assumption}\label{as.input}  The input $\u_t$ is an $\Fcal-$bandlimited graph process.
\end{assumption}}

Assumption~\ref{as.shift} focuses our attention to linear time-varying systems on graphs that are a function of the graph shift operator. In fact, for all network processes in Section~\ref{subsec.gsd_ssmg} this assumption holds. \rev{Assupmtion~\ref{as.input} requires the input signal to have a sparse GFT over time. That is, $\u_t$ should be a (piece-wise) smooth input signal on the graph, or have properties similar to the signals studied in \cite{hu2016msdGS,gomez2016netflix,yamanishi2004protein,wang2015local,narang2013localized,chen2015samp,marques2016sampling,tsitsvero2016signals,PDL2016_TSIPN,PDL2016_DLMS,didistributed,di2017adaptive,mei2015signal,LoukasPredict,Barbarossa2016Control,Gammacontrol}. From Definition~\ref{def.blimGP}, we should note that the bandlimitedness of $\x_t$ considers the sparsity in the GFT domain of all past realizations including those of ${\u_0, \ldots, \u_{t-1}}$. Furthermore, we will not consider Assumption~\ref{as.input} for the task of observability and will leverage it only for tracking.}

The subsequent proposition formalizes the above.
\rev{\begin{proposition}
Let $\x_t$ be a graph process that follows model \eqref{eq.state_space} and let Assumptions~\ref{as.shift} and \ref{as.input} hold. Then, $\x_t$ is an $\Fcal-$bandlimited graph process if and only if $\x_0$ is an $\Fcal-bandlimited$ graph signal.
\end{proposition}}
(The proof follows from simple algebra.)

With this in place, we can write the evolution of $\x_t$ as
\begin{subequations}
\label{eq.state_spec_gen}
\begin{align}
\label{eq.spec_state}
\tx_t &= \tA_{t-1}\tx_{t-1} + \tB_{t-1}\tilde{\u}_{t-1}\\
\y_{\Scal_t} &= \C_{\Scal_t}(\U_\Fcal\tx_t+ \v_t),
\label{eq.measurement_spec}
\end{align}
\end{subequations}
where $\tA_{t} = \U_\Fcal^{\herm}\A_{t}\U_\Fcal$ and $ \tB_{t} = \U_\Fcal^{\herm}\B_{t}\U_\Fcal$ are diagonal matrices containing the in-band spectrum of $\A_t$ and $\B_t$, respectively. 

Hereinafter, we will refer to systems of the form \eqref{eq.state_space} that can be written in the form \eqref{eq.state_spec_gen} as $\Fcal-$bandlimited systems on graphs. The $\Fcal-$bandlimited graph processes considered in this paper follow the evolutions~\eqref{eq.kf_equations} and \eqref{eq.spec_state} in the vertex and graph spectral domain, respectively. In the next section, we generalize condition \eqref{eq.rec_x} from the reconstruction of an $\Fcal-$ bandlimtied graph signal to the observation of an $\Fcal-$bandlimited graph process.

\section{Observing graph processes}
\label{sec.observ}

We start this section by adapting the definition of observability to our context \cite{Simon2006}.

\rev{\begin{definition}
An $\Fcal-$bandlimited system on graph is observable over the set $\Scal_{0:T} =\bigcup_{t = 0}^T\Scal_t = \{n \in \{1, \ldots, N\}; t \in \{0, \ldots, T\} |~c_{t,n} = 1	\}$ if for any $\Fcal-$bandlimited initial state $\x_0$ and some final time $T$, the initial state $\x_0$ can be uniquely determined in the absence of noise by the knowledge of the input $\u_t$ and measurement $\y_t$ for all $t \in \{0,\ldots, T\}$.
\end{definition}}

\noindent The set $\Scal_{0:T}$ specifies all graph-time locations where and when the nodes are sampled in the interval $\{0,\ldots, T\}$. Since for observability we need the knowledge of the input signal, Assumption~\ref{as.input} is not necessary here.

With the above formulation in place, we can answer the questions: $(Q1)$ \emph{Under which conditions is an $\Fcal-$bandlimited graph process observable from a few measurements?} $(Q2)$ \emph{When and where should we collect noisy measurements to estimate $\x_0$ up to a desired accuracy?}

To provide an answer, we write the relation between the measurement $\y_{t}$ and the initial $\Fcal-$bandlimited signal $\tx_0$ as
\begin{align}
\label{eq.obs_equations}
\begin{split}
\y_{t} &=\C_{\Scal_t}\U_\Fcal\tilde{\A}_{t,0}\tx_0 + \C_{\Scal_t}\U_\Fcal\sum_{\tau = 0}^{t-1}\tilde{\A}_{t,\tau+1} \tB_{\tau}\tu_\tau + \C_{\Scal_t}\v_t,
\end{split}
\end{align}\vspone\vspone\vspone
\noindent with \vspone\vspone\vspone\vspone\vspone
\begin{equation}\label{eq.expFmat}
    \tilde{\A}_{t,\tau}=\left\{
                \begin{array}{ll}
                  \tilde{\A}_{t-1}\tilde{\A}_{t-2}\ldots\tilde{\A}_{\tau},\quad t>\tau\\
                  \I_{|\Fcal|}, \quad t = \tau\\
                 \mathbf{0}_{|\Fcal|}\mathbf{0}_{|\Fcal|}\transp, \quad t < \tau.
                \end{array}
              \right.
\end{equation}
Let $\y_{0:T} = [\y_{0}^{\transp},\y_{1}^{\transp}, \ldots, \y_{T}^{\transp}]^{\transp}$ be the vector of measurements collected in the interval $\{0,\ldots, T\}$. Then, from \eqref{eq.obs_equations} we have
\begin{equation}
\label{eq.obs_compact}
\y_{{0:T}} = \O_{{0:T}}\tx_{0} + \J_{{0:T}}\u_{0:T-1} + \C_{\Scal_{0:T}}\v_{0:t},
\end{equation}
where
\begin{align}\label{eq.obsmat}
\begin{split}
\O_{{0:T}} \!&=\! [(\C_{\Scal_0}\U_\Fcal\tA_{0,0})^{\transp}\!\!,\! (\C_{\Scal_1}\U_\Fcal\tA_{1,0})^{\transp}\!\!, \ldots, (\C_{\Scal_T}\U_\Fcal\tA_{T,0})^{\transp}]^{\transp}\\
&=  \C_{\Scal_{0:T}}(\I_{T+1}\otimes\U_\Fcal)\tA_{0:T},
\end{split}
\end{align}
$\C_{\Scal_{0:T}} = \text{blkdiag}(\C_{\Scal_0}, \ldots, \C_{\Scal_T})$
$\tA_{0:T} = \left[\I_{|\Fcal|}, \tilde{\A}_{1,0}^{\transp}, \ldots, \tilde{\A}_{T,0}^{\transp}	\right]^{\transp}$, $\u_{0:T-1} = [\u_0^{\transp}, \u_1^{\transp}, \ldots, \u_{T-1}^{\transp}]^{\transp}$, and $\v_{0:T} = [\v_0^{\transp}, \v_1^{\transp}, \ldots, \v_T^{\transp}]^{\transp}$. $\J_{0:T}$ is the input evolution matrix in the interval $\{0,\ldots, T\}$ whose expression is not required for our derivations, but can be obtained from \eqref{eq.obs_equations}.

In the next section, we answer questions $(Q1)$ and $(Q2)$ for $\C_{{t}}$ being a deterministic sampler, while in Section~\ref{subsec.rand_obs} we consider the case where the entries of $\C_{{t}}$ follow a Bernoulli distribution.

\subsection{Observability with deterministic sampling}
\label{subsec.det_obs}

In this section, we consider the task of observability when the sampled nodes are chosen deterministically. Recall in this context that $\C_{\Scal_{0:T}}$ plays the role of the set projection matrix over the set ${\Scal_{0:T}}$.
Given then $\C_{\Scal_{0:T}}$, system \eqref{eq.state_spec_gen} is observable over ${\Scal_{0:T}}$ iff the observability matrix $\O_{0:T}$ in \eqref{eq.obsmat} is full rank \cite{Simon2006}, i.e., $\text{rank}(\O_{0:T}) = |\Fcal|$. Then, we have
\begin{equation}
\label{eq.obs_equations1}
\tx_0^o =\O_{0:T}^\dagger \left(\y_{0:T} - \J_{0:T}\u_{0:{T-1}}\right),
\end{equation}
which is also the least squares (LS) estimate of $\tx_0$ in the presence of noise $\v_t \neq 0$. From the structure of $\O_{0:T}$, a \emph{sufficient} condition for observability is that at least one of the block matrices $\C_{\Scal_t}\U_\Fcal\tF_{t,0}$ is of rank $|\Fcal|$, which requires at least $|\Fcal|$ nodes to be active for the specific $t$ (i.e., $|\Scal_t| \ge |\Fcal|$). This condition is similar to that of graph signal recovery via LMS \cite{PDL2016_TSIPN}, or RLS \cite{di2017adaptive} on graphs. However, while in adaptive graph signal recovery the goal is to reconstruct $\tx_0$ from multiple noisy realizations of the latter, here, we extend the recovery such that it encompasses also the model evolution (i.e., $\tA_{t,0}$) into the analysis. The latter allows to take measurements in a graph-time fashion, resulting in so-called \emph{graph-time} samples.
In this context, we claim the following.

\begin{proposition} 
\label{eq.prop_det_observability}
An $\Fcal-$bandlimited system on graph is observable over the set $\Scal_{0:T}$ only if at least $|\Fcal|$ graph-time samples are taken in the time interval $\{0, \ldots, T\}$. These samples can be taken by $|\Fcal|$ nodes at a fixed time instant, by one node in $|\Fcal|$ time instants, or a combination of the two.
\end{proposition}

Put simply, the condition in Proposition \ref{eq.prop_det_observability} is equivalent to
\begin{equation}
\label{eq.obs_cond}
|\Scal_{0:T}| \ge |\Fcal|,
\end{equation}
i.e., the cardinality of the sampling set must be greater than or equal to the process bandwidth. However, \eqref{eq.obs_cond} is only a necessary condition for observability. In fact, $\O_{0:T}$ may be easily ill-conditioned depending on the particular location of these samples {and the spectral support of $\tx_0$.}

\rev{It is then paramount to carefully pick the samples in a graph-time fashion such that $\O_{{0:T}}$ is of full rank $|\Fcal|$, and in the presence of noise $\v_t \neq 0$, possibly also well-conditioned.} Put differently, the sampling set should satisfy
\begin{equation}
\label{eq.ranktF1}
\text{rank}\Big(\sum_{t=0}^T\tA_{t,0}^\herm\U_\Fcal^\herm\C_{\Scal_t}\U_\Fcal\tA_{t,0} \Big) = |\Fcal|,
\end{equation}
where the single shot graph signal reconstruction \cite{tsitsvero2016signals} is the special case $T = 0$.
The following theorem generalizes \eqref{eq.rec_x} to a necessary and sufficient condition for the observability of an $\Fcal$-bandlimited graph process over a sampling set.
\rev{\begin{theorem}
\label{theo_obs}
An $\Fcal-$bandlimited system on graph is observable over the set $\Scal_{0:T}$ if and only if
\begin{equation}
\label{eq.theo2}
\|\C_{\Scal_{0:T}^c}(\I_{T+1}\!\otimes\!\U_\Fcal)\| < \frac{s^2_{\text{min}}(\tA_{0:T})}{s^2_{\text{max}}(\tA_{0:T})},
\end{equation}
where $\C_{\Scal_{0:T}^c} = \I_{N(T+1)} - \C_{\Scal_{0:T}}$ is the operator that projects onto the complementary set $\Scal_{0:T}^c =  \{n \in \{1, \ldots, N\}; t \in \{0, \ldots, T\} ~|~ c_{t,n} = 0	\}$ and $s_{\text{min}}(\tA_{0:T})$, $s_{\text{max}}(\tA_{0:T})$ indicate the minimum and maximum singular values of $\tA_{0:T}$, respectively.
\end{theorem}}
Condition \eqref{eq.theo2} (analogous to \eqref{eq.rec_x}) is again related to the localization properties of graph signals involving also the evolution model of the latter. It implies that in their \emph{evolution} there are no $\Fcal$-bandlimited graph processes perfectly localized on the complementary set $\Scal_{0:T}^c$. {The single shot condition \eqref{eq.rec_x} is obtained for $T = 0$.}

We conclude this part with the following observation.

\begin{remark}
While \eqref{eq.obs_equations1} is an option to estimate $\x_0$ in the LS sense, one can also rely on the time-invariant results by considering only one realization $\y_{t}$. In this case, the presence of $\tA_{t,0}$ should also be considered. Thus, when rank$(\tA_{t,0}) < |\Fcal|$, the recovery over singular observations is not possible. In a time-varying fashion, we exploit the successive realizations for estimating $\x_0$. Furthermore, since we must deal with noise, operating in a graph-time fashion makes the recovery more robust to bad noise realizations for a particular $t$.
\end{remark}

\textbf{MSE analysis.} We here quantify how the sampling set $\Scal_{0:T}$ affects the MSE of the LS estimate \eqref{eq.obs_equations1}. The latter will be then used as a criterion to collect the graph-time samples. The main result is given by the following proposition.

\begin{proposition}
\label{theo_mse_det_obs} Given an $\Fcal-$bandlimited graph process following the model \eqref{eq.state_spec_gen} and assuming the result of Theorem~\ref{theo_obs} holds. Then, the \MSE of the \LS observed signal $\tx_0^o$ is 
\begin{align}
\label{mse_theo1}
\begin{split}
\MSE \!&=  \Exp\left\{\|\tx_0^o - \tx_0	\|^2	\right\} =  \Exp \left\{\tr\left[	(\tx_0^o - \tx_0)(\tx_0^o - \tx_0)^{\herm}	\right]	\right\}\\
&= \!\sigma_v^2\tr\!\left\{\!\left[\!\tA_{0:T}^{\herm}(\I_{T+1}\! \otimes\! \U_\Fcal)^{\herm}\C_{\Scal_{0:T}}(\I_{T+1}\! \otimes\! \U_\Fcal)\tA_{0:T}\!	\right]^{-1}	\right\}\!\!.
\end{split}
\end{align}
\end{proposition}

(The claim follows from the covariance matrix of the LS estimator \cite{kay2013fundamentals}.)

Besides characterizing the impact of the graph-time samples on the MSE\footnote{The absence of model noise in \eqref{eq.kf_equations} allows us to find a closed-form expression for the MSE, rather than an upper bound. Moreover, it matches perfectly models \eqref{eq.inst_diff}-\eqref{eq.ss_solarma1}.}, expression \eqref{mse_theo1} shows that not only the number of selected samples plays a role, but also their location in graph and time. In the sequel, we show how to select these samples such that a target MSE \eqref{mse_theo1} is guaranteed.

\textbf{Sampling strategy.} Given \eqref{mse_theo1}, we follow a sparse sensing approach \cite{joshi2009, chepuri2016sparse} to design the sampling set $\Scal_{0:T}$ such that a target MSE estimation performance is guaranteed. The latter is achieved as the solution of the convex problem
\begin{equation}\label{opt.mse_sel}
\begin{aligned}
& \underset{\c_{0:T}}{\text{minimize}}
& & \mathbf{1}^{\transp}_{N\times(T+1)}\c_{0:T} \\
& \text{subject to}
& & \tr\left[\left(\bPsi_{0:T}^{\herm}\C_{\Scal_{0:T}}\bPsi_{0:T}	\right)^{-1}	\right] \le \frac{\gamma}{\sigma_{v}^2},\\
&&& \C_{\Scal_{0:T}} = \text{diag}(\c_{0:T}),\\
&&& \bPsi_{0:T} = (\I_{T+1}\! \otimes\! \U_\Fcal)\tA_{0:T},\\
& & & 0 \le c_{0:T,i} \le 1,\quad i = 1, \ldots, N(T+1),
\end{aligned}
\end{equation}
where the objective function is the $l_1$-surrogate of the $l_0$-norm and imposes sparsity in $\Scal_{0:T}$; the constant $\gamma > 0$ imposes a target MSE performance; and the last constraint is the relaxation of the Boolean constraint $c_{0:T,i} \in \left\{0,1	\right\}$ to the box one\footnote{In a second step, randomized rounding or thresholding can be used to project the optimal solution $\c_{0:T}^*$ of \eqref{opt.mse_sel} to the $\{0,1	\}^{N(T+1)}$ space \cite{joshi2009}.}. Alternatively, one can adopt a greedy approach similar to \cite{chamon2018greedy} for building $\Scal_{0:T}$. Obviously, we can also consider the opposite problem where the aim is to minimize the MSE, while imposing a fixed budget on the selected number of samples. The latter translates as well into a convex problem.


\subsection{Observability with random sampling}
\label{subsec.rand_obs}

In this section, we consider the case where the entries of $\C_{\Scal_{t}}$ in \eqref{eq.measurement} are i.i.d. in time Bernoulli random variables with expected value $\bC = \text{diag}(\bar{\c})$. 
\rev{Let then $\bar{\Scal} = \{n \in \{1, \ldots, N\}|\bar{c}_n >0\}$ be the expected sampling set, i.e., the set of nodes that are sampled with a probability greater than zero.}
The task here is to answer questions $(Q1)$ and $(Q2)$ w.r.t. $\bar{\Scal}$. As we show at the end of this section, one major benefit of this approach is that the sparse sensing design of $\bar{\Scal}$ avoids the relaxation techniques used in \eqref{opt.mse_sel}.

Given the measurements in the interval $\{0, \ldots, T\}$ \eqref{eq.obs_compact}, for a realization of $\C_{\Scal_{0:T}}$, we define
\begin{equation}
\label{eq.mean_vChange}
\z_{0:T} = \y_{0:T} - {\J}_{0:T}\u_{0:T} = \O_{{0:T}}\tx_{0} + \C_{\Scal_{0:T}}\v_{0:T},
\end{equation}
i.e., we subtract each realization\footnote{This is analogous to our graph deterministic observability, or observability in linear systems, where the realizations of the input signal should be known. This is not a problem since the realization of $\C_{\Scal_{0:T}}$ is known.} of the input signal before analyzing the observability properties. From \eqref{eq.obs_cond}, a necessary condition for the instantaneous observability matrix $\O_{{0:T}}$ to be full rank is that the instantaneous sampling set ${\Scal_{0:T}}$ in $\{0, \ldots, T\}$ has a cardinality greater than, or equal to, the signal bandwidth. 
Given the structure of $\O_{{0:T}}$ ($\C_{\Scal_{0:T}}$) in \eqref{eq.obsmat}, it is obvious that rank$(\O_{{0:T}})$ \big(rank($\C_{\Scal_{0:T}}$)\big) depends on the rank$(\bC)$, i.e., on the cardinality of the nodes that are sampled with a probability strictly greater than zero. The subsequent proposition formalizes the above as a necessary condition.

\rev{\begin{proposition}
\label{prop.obs_mean} 
Consider an $\Fcal-$bandlimited system on graph and given the diagonal sampling matrix $\C_{\Scal_{t}}$ with i.i.d. in time Bernoulli entries and expected value $\bC$. A necessary condition for the observability of the system from samples taken randomly in the interval $\{0, \ldots, T\}$ is that at least $\lceil|\Fcal|/(T+1) \rceil$ nodes are sampled with a probability greater than zero.
\end{proposition}}
%
%
\rev{That is, differently from the deterministic node sampling, the observability of a graph process is now related to the expected sampling set $\bar{\Scal}$. Put simply, the constraint in Proposition~\ref{prop.obs_mean} is equivalent to
\begin{equation}
|\bar{\Scal}| \ge \lceil|\Fcal|/(T+1) \rceil.
\end{equation}
It must be noted that for $T \ge |\Fcal|$ there is the potential to observe an $\Fcal-$bandlimited graph process by allowing only one node to randomly take measurements.} The above result, though novel from the random sampling viewpoint, is not entirely surprising. In fact, in \cite{marques2016sampling} it has been seen that a graph signal can be reconstructed also by sampling successive aggregations of a single node. Hence, by bringing the time into the play, one node {can collect different linearly independent measurements in time} and will be able to observe the process.


\rev{However, since the node sampling is random in a finite interval $\{0, \ldots, T\}$, there is always a possibility that the instantaneous sampling set ${\Scal_{0:T}}$ has a cardinality smaller than $|\Fcal|$. The following corollary quantifies the latter.
\begin{corollary}\label{eq.corr_prob}
Given the sampling matrix $\C_{\Scal_t}$ in \eqref{eq.state_space} with i.i.d. in time Bernoulli entries and expected value $\bC$. The probability that the cardinality of the instantaneous sampling set $\Scal_{0:T}$ is smaller than the process bandwidth $|\Fcal|$ is
\begin{equation}\label{eq.prob}
\textnormal{Pr}\big(|\Scal_{0:T}| < |\Fcal|	\big) = \sum_{k = 0}^{|\Fcal|-1}\frac{\alpha^ke^{-\alpha}}{k!},
\end{equation}
where $\alpha = (T+1)\1_N\transp\bar{\c}$ is the mean of the Poisson distribution.
\end{corollary}
\noindent This probability drops to zero even for moderate values $\bar{\c}$ as long as $N$ is of the order of $100$ nodes and $T$ is relatively large. In Section~\ref{subsec.num_obs}, we show with real data that this probability drops below machine precision. To further quantify the impact of $\bC = \text{diag}({\bar{\c}})$ on the process observability, we perform next an MSE analysis of the LS estimated state.}

\textbf{MSE analysis.} To render a MSE analysis tractable, we follow a similar procedure as used for the Cram\'{e}r-Rao lower bound\footnote{{Since the measurements are the product of a Bernoulli and a Gaussian random variable, the joint pdf does not satisfy the CRLB regularity condition.}} (CRLB) \cite{kay2013fundamentals}, which quantifies the lowest MSE estimate $\tx_0^o =  \O_{0:T}^\dagger\z_{0:T}$. The following proposition quantifies this finding.
\begin{proposition}
\label{prop_mseRndObs}
Given an $\Fcal-$bandlimited graph process following model \eqref{eq.state_spec_gen} and given $\C_{\Scal_t}$ a diagonal sampling matrix with i.i.d. in time Bernoulli entries and expected value $\bC$. The \MSE of the \LS observed signal $\tx_0^o =  \O_{0:T}^\dagger\z_{0:T}$ is then lower-bounded by
\begin{align}
\label{eq.propMSE_eq2}
\begin{split}
\MSE \ge \sigma_v^2\tr\left\{\left[\tA_{0:T}^\herm\left(	\I_{T+1}\otimes\U_\Fcal^\herm\bC\U_\Fcal	\right)\tA_{0:T}\right]^{-1}\right\}.
\end{split}
\end{align}
\end{proposition}

{Besides providing a statistical measure of the lowest achievable MSE for a particular $\bC$, the lower bound \eqref{eq.propMSE_eq2} can also be used as a design criterion to find these sampling probabilities. This aspect is covered in more detail next.
}

\textbf{Sampling strategy.} Following the same principle as in \cite{chepuri2016sparse}, we design the expected sampling set $\bar{\Scal}$ in a sparse sensing fashion, where instead of using the CRLB as a design criterion, we consider the lower bound \eqref{eq.propMSE_eq2}. Then, $\bar{\c}$ and therefore $\bar{\Scal}$ are found as the solution of the convex problem
%
%
%
{
\begin{equation}
\label{eq.opt_prob_obs}
\begin{aligned}
& \underset{\bar{\c}, \mathbold{\gamma} \in \mathbb{R}^{|\Fcal|}}{\text{minimize}}
& & \mathbf{1}^{\transp}_N\bar{\c} \\
& \text{subject to}
& & \tr\left\{\left[\tA_{0:T}^\herm\left(	\I_{T+1}\otimes\U_\Fcal^\herm\bC\U_\Fcal	\right)\tA_{0:T}\right]^{-1}\right\} \le \frac{\gamma}{\sigma_v^2},\\
& & & \bC = \text{diag}(\bar{\c}), \\
& & & c_{\text{min}} \le \bar{c}_{n} \le c_{\text{max}} ,\quad n = 1, 2, \ldots, N,\\
& & & 0 \le c_{\text{min}} \le c_{\text{max}} \le 1. 
\end{aligned}
\end{equation}
}\vsp\vsp\vsp

{
Even though conceptually equivalent to \eqref{opt.mse_sel}, problem \eqref{eq.opt_prob_obs} differs in two main aspects, which preserve the optimality of the solution. First, the objective function is not a surrogate anymore of the $l_0$-norm. Rather, it is the true function (i.e., the overall sampling rate) that we want to minimize. Second, the convex box constraint $c_{\text{min}} \le \bar{c}_{n} \le c_{\text{max}}$ is not a relaxation anymore, since now we directly optimize over the sampling probabilities for some rate allocation bounds $c_{\text{min}}$ and $c_{\text{max}}$. In \eqref{eq.opt_prob_obs}, one can add also a constraint on the probability criterion \eqref{eq.prob}. However, the latter should be upper-bounded since it is not a convex function. As we show in Section~\ref{sec.simulations} with the ETEX dataset, the latter is not necessary since a small enough $\gamma$ on the MSE will trade well the sampling probabilities with the performance.
}
%
\vsp\vsp\vsp

\section{Tracking graph processes}
\label{sec.kalman}

We now consider the task of tracking a bandlimited graph process from a subset of nodes chosen deterministically. We make use of the Kalman filter, which matches perfectly the system \eqref{eq.state_spec_gen}. First, we introduce the KF algorithm for time-varying scenarios and provide conditions on the sampling set to optimally track the graph process. Then, we show how the proposed KF specializes for time-invariant models (as the ones in Section~\ref{subsec.gsd_ssmg}) and provide conditions for the sampling set to ensure a steady-state performance. For both cases, we provide sampling strategies for designing the sampling set with given tracking guarantees.
\vsp\vsp\vsp

\subsection{Kalman filtering for time-varying models}
\label{subs.KF}

Following \cite{Simon2006}, for the $\Fcal-$bandlimited system \eqref{eq.state_spec_gen}, the KF on graphs evolves as described in Algorithm~\ref{alg_kf}. It initializes the \emph{a posteriori} state estimate $\tx_0^+$ to a random vector and the \emph{a posteriori} error covariance matrix $\P_0^+$ to a scaled identity. The update of $\P_t^-$ in step  $ii)$ accounts also for the state model noise $\tw_t = \U_\Fcal\w_t$, which is considered zero-mean with covariance matrix $\tilde{\bSigma}_w= \U_\Fcal^\herm\bSigma_w\U_\Fcal$ and independent from $\v_t$.

The Kalman gain matrix $\K_t$ computed in step $iii)$ leads to the minimum average \emph{a posteriori} MSE, i.e., $\tr(\P_t^+)$. From its expression, it is clear that $\K_t$ (and thus $\P_t^+$) is highly correlated with the sampling set at time $t$, i.e., $\C_{\Scal_t}$. In fact, for rank$(\C_{\Scal_t}) = R < |\Fcal|$ and assuming rank$( \P_t^-) = |\Fcal|$, then rank$(\C_{\Scal_t}\U_\Fcal\P_t^-\U_\Fcal^\herm\C_{\Scal_t} + \C_{\Scal_t}\bSigma_v\C_{\Scal_t}) = R$. As a consequence rank$(\K_t) \le R < |\Fcal|$. Thus, the $|\Fcal| \times N$ Kalman gain matrix $\K_t$ can be full rank only if

\begin{equation}
\label{eq.nec_KF}
|\Scal_t| \ge |\Fcal|.
\end{equation}
That is, KF on graphs will fully exploit its Kalman gain (and thus track better) only if the number of sampled nodes \emph{for each} $t$ is greater than or equal to the signal bandwidth. The necessary condition \eqref{eq.nec_KF} extends condition \eqref{eq.obs_cond} from the observability of a bandlimited graph process in an interval to the tracking task.

The impact of the sampled nodes $\C_{\Scal_t}$ in KF is highlighted in $\P_t^+$ (e.g., step v) in Algorithm~\ref{alg_kf}). In the sequel, we will exploit this benefit to design $\Scal_t$ such that a target \emph{a posteriori} MSE estimation accuracy is guaranteed. 

\begin{algorithm}[t]
\caption{\textbf{:~Kalman filtering on graphs}}
\vspace{.1cm}
Initialize $\tx_0^+$ and $\P_0^+$. For $t>0$ repeat:\\
 i) Update the \emph{a priori} state estimate $\tx_t^-$ as:
 \begin{equation*}\label{eq.prior_Kf_x}
 \tx_t^- = \tA_{t-1}\tx_{t-1}^{+}+ \tB_{t-1}\tu_{t-1};
 \end{equation*}
 ii) Update the \emph{a priori} error covariance matrix $\P_t^{-}$ as:
  \begin{equation*}\label{eq.prior_Kf_cov}
 \P_t^{-} = \tF_{t-1}\P_{t-1}^+\tF_{t-1}^{\herm} + \tilde{\bSigma}_w;
 \end{equation*}
 iii) Compute the Kalman gain matrix $\K_t $ as:
   \begin{equation*}\label{eq.Kalman_gainmatrix}
\K_t = \P_t^-\U_\Fcal^{\herm}\C_{\Scal_t}\left(\C_{\Scal_t}\U_\Fcal\P_t^-\U_\Fcal^{\herm}\C_{\Scal_t} + \C_{\Scal_t}\bSigma_v\C_{\Scal_t}	\right)^{\dagger};
 \end{equation*}
 iv) Update the \emph{a posteriori} state estimate $ \tx_t^+$ as:
    \begin{equation*}\label{eq.Kalman_gainmatrix}
 \tx_t^+ = \tx_t^- + \K_t\left(\y_t - \C_{\Scal_t}\U_\Fcal\tx_t^-		\right);
 \end{equation*}
 v) Update the \emph{a posteriori} error covariance matrix $\P_t^{+}$ as:
 \begin{align*}
\begin{split}
\P_t^+ &= \P_t^- - \P_t^-\U_\Fcal^{\herm}\C_{\Scal_t}\K_t^\herm - \K_t\C_{\Scal_t}\U_\Fcal\P_t^- \\
&\quad + \K_t\C_{\Scal_t}\bSigma_v\K_t^{\herm}  +\K_t\C_{\Scal_t}\U_\Fcal\P_t^-\U_\Fcal^{\herm}\C_{\Scal_t}\K_t^{\herm}.
\end{split}
\end{align*}
\label{alg_kf}
\end{algorithm}

\textbf{Sampling strategy.} Condition \eqref{eq.nec_KF} suggests that there is a minimum number of nodes, tightly related to the signal bandwidth, that must be sampled to fully exploit the Kalman gain matrix. However, from the expression of $\P_t^+$, it is once again clear that their location in the graph is as important as $|\Scal_t|$ to achieve a good tracking performance.

To optimally select the sampled nodes, we adopt a sparse sensing approach that involves the posterior CRB (PCRB) for the state estimation $\tx_t$ \cite{chepuri2016sparse,zuo2007posterior,tichavsky1998posterior}. Given the log-likelihood of the measurements satisfies the regularity condition $\Exp[\del\text{ln}p(\y_t; \tx_t)/\del\tx_t] = \mathbf{0}_N$, $\forall t$, the PCRB satisfies
\begin{equation}
\label{eq.pcrb1}
\P_t^+ = \Exp\left[\left(\tx_t^+-\tx_t\right)\left(\tx_t^+-\tx_t\right)^\herm	\right]\ge \F_t^{-1}(\tx_t),
\end{equation}
where $\F_t(\tx_t)$ is the posterior Fisher information matrix. 

For linear systems in additive Gaussian noise, as the one considered in this work, relation \eqref{eq.pcrb1} holds with equality (i.e., the Kalman filter is optimal) and is independent of $\tx_t$.
%
%
This suggests that designing the sampling set w.r.t. $\F_t(\tx_t)$ leads to the same result as working with the \emph{a posteriori} error covariance matrix $\P_t^+$. For the KF in Algorithm~\ref{alg_kf} the posterior FIM is \cite{tichavsky1998posterior}:
\begin{equation}
\label{eq.fim_update}
\F_t(\tx_t) = \left(\tA_t\F_{t-1}^{-1}(\tx_t)\tA_t^{\transp} + \bSigma_{\tilde{w}} 	\right)^{-1} +\sum_{n = 1}^Nc_{t,n}\F_{t,n}^o(\tx_t),
\end{equation}
where $c_{t,n}$ is the $n$th diagonal entry of $\C_{\Scal_t}$, and $\F_{t,n}^o(\tx_t) = \sigma^{-2}_v{c_{t,n}}\u_{\Fcal,n}\u_{\Fcal,n}^{\transp}$ is the FIM related to the $n$th node observation at time $t$. 
%
%
The first term in \eqref{eq.fim_update} denotes the prior FIM related to the tracking history up to $t-1$. 

By substituting the expression for $\F_{t,n}^o(\tx_t)$ and rearranging the sum, we obtain the \emph{a posteriori} FIM
\begin{equation}
\label{eq.fim_update1}
\F_t(\tx_t) = \left(\tA_t\F_{t-1}^{-1}(\tx_t)\tA_t^{\transp} + \bSigma_{\tilde{w}} 	\right)^{-1} + \U_\Fcal^{\herm}\C_{\Scal_t}\bSigma_v^{-1}\U_\Fcal.
\end{equation}
Following once again the sparse sensing idea, the instantaneous sampling set $\Scal_t$ can be built by solving
\begin{equation}
\label{eq.KF_nodeS}
\begin{aligned}
& \underset{{\c_t}}{\text{minimize}}
& & \mathbf{1}^{\transp}{\c_t} \\
& \text{subject to}
&& \F_t(\tx_t) \ge \gamma\I_{|\Fcal|},\\
& & & \C_{\Scal_t} = \text{diag}(\c_t),\\
& & & \mathbf{0}_N \le \c_t \le \mathbf{1}_N.
\end{aligned}
\end{equation}
Problem \eqref{eq.KF_nodeS} generalizes \eqref{opt.mse_sel} to the time-varying case, where all the remarks about the optimallity of the solution extend also here.
%
\begin{remark}
From a practical viewpoint, the presented KF approach presents two main challenges in large graphs. {First, the computation $\K_t$ involves the pseudo-inverse of an $|\Fcal| \times N$ matrix with at most $|\Scal_t|^2$ (by construction) non zero elements. The latter may result computationally prohibitive for $|\Scal_t| \to N$.} This issue can be easily addressed with the sequential implementation of the Kalman filter \cite{Simon2006}. 
Second, the node selection strategy involves solving for each time instant $t$ an SDP problem, which for large $N$ may result in prohibited costs \cite{vandenberghe1996semidefinite}. The latter issue can be addressed with greedy solutions such as \cite{krause2008optimizing}.
\end{remark}
\vspone\vspone\vspone\vspone\vspone
\subsection{Steady-state Kalman filtering on graphs}
\label{subsec_ssKF}

We focus here on a time-invariant $\Fcal-$bandlimited system on graphs, which specializes the above derivations to models \eqref{eq.inst_diff}-\eqref{eq.ss_solarma1}. These systems often lead to a convergent state, which can be exploited to design a fixed sampling set for all $t$, i.e., $\C_{\Scal_t} = \C_\Scal$ $\forall t$ with given steady-state performance guarantees. 

From \cite{Simon2006}, the \emph{a priori} error covariance matrix $\P_t^-$ will converge to the unique limit $\P_\infty$ if:

$i)$ the pair $(\tA,\tB)$ is stabilizable;

$ii)$ the pair $(\tA,\C_\Scal\U_\Fcal)$ is detectable. 

\noindent The first condition is a characteristic of the graph process and is application specific. However, for stable time-invariant graph processes (as the one of interest in this section) this condition is satisfied. The second condition restricts the sampled nodes and their location in the graph to guarantee the steady-state convergence. From linear systems theory, a useful result is that an observable system is also detectable. Thus, by exploiting the findings in Section~\ref{subsec.det_obs}, the KF on graphs is convergent if the limiting observability matrix
\begin{equation}\label{eq.lim_obsM}
\O_\infty = \lim_{T\to\infty}\O_{0:T} = \lim_{T\to\infty}(\I_{T+1}\otimes\C_\Scal\U_\Fcal)\tA_{0,T}
\end{equation}
is full rank, or with similar arguments as in Theorem~\ref{theo_obs} if
\begin{equation}\label{eq.lim_obsM1}
\lim_{T\to\infty}\|\I_{T+1} \otimes \U_\Fcal^{\transp}\D_{\Scal^c}\U_\Fcal\| \le \lim_{T \to \infty}\frac{s^2_{\text{min}}(\tA_{0:T})}{s^2_{\text{max}}(\tA_{0:T})},
\end{equation}
where $\Scal^c = \mathcal{V}/\Scal$ denotes again the complementary sampling set. Conditions \eqref{eq.lim_obsM}, once again relates the complementary sampling set with the localization properties of the graph process throughout its temporal evolution (though in practice it is sufficient to hold for $T \gg 0$).

For a fixed $\Scal$, the \emph{a priori} error covariance matrix satisfies the discrete algebraic Riccatti equation (DARE)
\begin{align}\label{eq.dare}
\begin{split}
&\P_\infty = \tA\P_\infty\tA^{\transp} + \tilde{\bSigma}_w - \tA\P_\infty\U_\Fcal^{\transp}\C_\Scal \times\\
&\left(\!\C_\Scal\U_\Fcal\P_\infty\U_\Fcal^{\transp}\C_\Scal \!+\! \C_\Scal\bSigma_v\C_\Scal		\!\right)^{\!\dagger}\!\!\!\C_\Scal\U_\Fcal\P_\infty\tA^{\transp} .
\end{split}
\end{align}
Consequently, the steady-state Kalman gain matrix\footnote{Despite not having a closed form solution, the DARE equation \eqref{eq.dare} admits a numerical solution \cite{Simon2006}.} is
\begin{equation}\label{eq.ssKG}
\K_\infty = \P_\infty\U_\Fcal^{\transp}\C_\Scal\left(\C_\Scal\U_\Fcal\P_\infty\U_\Fcal^{\transp}\C_\Scal + \C_\Scal\bSigma_v\C_\Scal	\right)^\dagger,
\end{equation}
with posterior state estimate
\begin{equation}\label{eq.ssKG1}
\tx_t^+ = \left(\I_{|\Fcal|} - \K_\infty\C_\Scal\U_\Fcal	\right)\tA\tx_{t-1}^+ + \K_\infty\y_t.
\end{equation}

Differently from the time-varying scenario, the above steady-state KF is only asymptotically optimal. To avoid the matrix inversion in \eqref{eq.ssKG}, we can rely once again on the sequential implementation \cite{Simon2006}. From the expression of $\K_\infty$, a necessary condition to fully exploit the steady-state Kalman gain matrix is that the cardinality of the sampling set should satisfy $|\Scal| > |\Fcal|$. That is, the graph structure, the process bandwidth, and the cardinality of the sampling set are once again tightly related to fully exploit the KF benefits in ensuring a predefined performance.

\textbf{Sampling strategy.} 
Similar to the previous selection strategies, the optimal sampling set that minimizes the steady-state performance for a fixed number of available nodes is found as
\begin{equation}
\label{eq.KF_node_ss}
\begin{aligned}
& \underset{{\c}}{\text{minimize}}
& & \tr(\P_\infty) \\
& \text{subject to}
& &\C_\Scal = \text{diag}(\c),\\
&&& \|\c\|_0 = |\Scal|,\\
&&&\c \in \{0,1\}^N.
\end{aligned}
\end{equation}
Problem~\eqref{eq.KF_node_ss} provides the optimal solution for the sampling set that guarantees the best steady-state estimation accuracy. However, even by relaxing the non convex constraint as in \eqref{opt.mse_sel} and \eqref{eq.KF_nodeS}, the impossibility of having a closed form solution for the DARE \eqref{eq.dare} renders \eqref{eq.KF_node_ss} intractable. This result is not entirely surprizing, since the latter issue is commonly present in the sensor selection literature \cite{dhingra2014admm,zhang2017sensor,gupta2006stochastic}.

A common way to tackle problem \eqref{eq.KF_node_ss} is by greedy algorithms \cite{yang2015deterministic,tzoumas2016sensor,zhang2017sensor}. For our specific case, we adopt the strategy from \cite{zhang2017sensor}, where the node sampling proceeds as described in Algorithm~\ref{alg_greedyssKF}. The sampling strategy considers starting with an empty sampling set and greedily adding the nodes that give the smallest increment in the steady-state estimation error (e.g., step v)). The solution of DARE $\P_\infty(\Scal \cup \{n\})$ in step iii) considers solving numerically \eqref{eq.dare} for $\Scal = \Scal \cup \{n\}$. Finally the algorithm stops when the desired cardinality of $\Scal$ is achieved.

\begin{algorithm}[t]
\caption{\textbf{:~Greedy node sampling algorithm from \cite{zhang2017sensor} for problem \eqref{eq.KF_node_ss}}}
\vspace{.1cm}
Start with an empty sampling set $\Scal = \emptyset$, a fixed cardinality $|\Scal|$ and counter $c = 0$\\
 i) \textbf{FOR} $c \le |\Scal|$\\
 ii)\qquad \textbf{WHILE} $n \in \Scal^c$\\
 iii) \qquad\quad Compute $\tr(\P_\infty(\Scal \cup \{n\}))$ in \eqref{eq.dare};\\
 iv)\qquad \textbf{END FOR}\\
 v)\qquad ~Select $n$ as {argmin}$_n\tr(\P_{\infty}(\Scal \cup \{n\}))$;\\
 vi)\qquad Update the sampling set $\Scal = \Scal \cup \{n\}$;\\
 vii)\qquad \hskip-1mmUpdate the counter $c = c + 1$;\\
 viii)\textbf{END WHILE}
\label{alg_greedyssKF}
\end{algorithm}

For the steady-state KF on graphs \eqref{eq.dare}-\eqref{eq.ssKG1}, the greedy Algorithm~\ref{alg_greedyssKF} is optimal with respect to problem \eqref{eq.KF_node_ss} if \cite{zhang2017sensor}:

$i)$ the measurement noise $\v_t$ is uncorrelated;

$ii)$ the set of sensor information matrices $\{\F_1, \ldots, \F_N\}$ with $\F_n = \sigma_v^{-2}\u_{\Fcal,n}\u_{\Fcal,n}^{\transp}$ is totally ordered w.r.t. the order relation of positive semidefiniteness.


\noindent The first condition is easily met in practice. The second condition relates the graph topology and the graph signal bandwidth with the optimal sampling set. It implies that for two different sets $\Scal^\prime$ and $\Scal^{\prime\prime}$ with $\F(\Scal) = \sum_{n = 1}^{|\Scal|}\sigma_v^{-2}\u_{\Fcal,n}\u_{\Fcal,n}^{\transp} = \U_\Fcal^{\transp}\C_\Scal\bSigma_v^{-1}\U_\Fcal$, if $\F(\Scal^\prime) \succeq \F(\Scal^{\prime\prime})$ it holds that $\tr(\P_\infty(\Scal^\prime)) \le \tr(\P_\infty(\Scal^{\prime\prime}))$. Thus the node sampled in the set $\Scal^{\prime}$ is a better choice than the node sampled in the set $\Scal^{\prime\prime}$ \cite{zhang2017sensor}. 

As we illustrate in Section~\ref{subsec.num_obs}, the proposed KF on graph optimally tracks graph processes and outperforms other alternatives in terms of estimation accuracy.

\vsp\vsp\vsp
\section{Numerical evaluation}
\label{sec.simulations}
We now corroborate our findings with numerical results using both synthetic and real data. We start with the task of observability and then we move to the KF. In these simulations, we made use of the GSP box \cite{perraudin2014gspbox} and CVX \cite{grant2008cvx}.\vspone\vspone\vspone

\subsection{Observing graph processes}
\label{subsec.num_obs} 

We first test the observability with deterministic sampling on the Molene weather data set\footnote{Data publicly available at \texttt{https://donneespubliques.\\meteofrance.fr/donnees\_libres/Hackathon/RADOMEH.tar.gz}.} and then the observability with random sampling on the European tracer experiment (ETEX) data set\footnote{Data publicly available at https://rem.jrc.ec.europa.eu/RemWeb/etex/ .} \cite{nodop1998field}.

\textbf{Obs. with deterministic sampling.} The Molene weather data set consists of $R = 744$ hourly temperature recordings collected in January 2014 over 32 cities in the region of Brest, France. {The graph is a $k$-nearest neighbour ($k$NN) \cite{perraudin2014gspbox} graph with $k = 3$.} We consider a single recording\footnote{The graph signal consists of the measured temperature after subtracting their average value.} and then diffuse it following model \eqref{eq.inst_diff} with $w = 1.5$ and $T = 10$. 

\begin{figure}[t]
  \centering
    \includegraphics[width=0.5\textwidth, trim={.1cm 0 .25cm .3cm},clip]{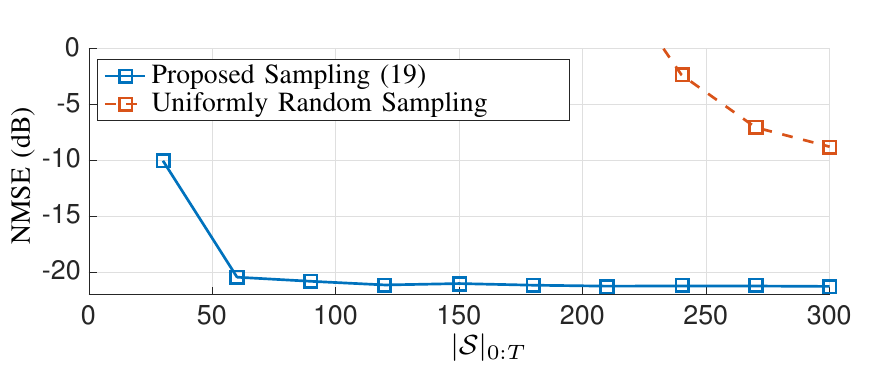}
    \vskip -.45cm
      \caption{NMSE versus the number of samples for the sampling algorithm \eqref{opt.mse_sel} and uniformly random sampling. The graph process has a perfectly localized spectrum on $|\Fcal| = N = 32$, i.e., the entire bandwidth.}
          \label{fig.detObs1}
\end{figure}

\setlength{\extrarowheight}{2.0pt}
\setlength{\tabcolsep}{9pt}
\begin{table}
\linespread{1.5}
\centering
\caption{ Theoretical, empirical NMSE, and the cardinality of the sampling set $\Scal_{0:T}$ for different values of $\gamma$ in \eqref{opt.mse_sel}.
}\vskip-5mm
\label{table:coeff}
\begin{center}
\vspace{-1mm}
 \begin{tabular}{@{}c @{}c @{}c @{}c @{}c@{}} \toprule
\textit{\ \ } & $\gamma = 2.05$ & $\gamma = 2.5$& $\gamma = 3$& $\gamma = 3.5$ \\  
 \midrule 
Theo. \eqref{mse_theo1} & \hspace{3mm}\begin{tabular}{@{}c@{}} $-21.26$dB\end{tabular}  & \hspace{3mm}\begin{tabular}{@{}c@{}}  $-20.42$dB\end{tabular}  & \hspace{3mm}\begin{tabular}{@{}c@{}}  $-19.64$dB\end{tabular}& \hspace{3mm}\begin{tabular}{@{}c@{}}  $-19.32$dB\end{tabular} \\ [0.0ex]
\rowcolor{black!7}[0pt][0pt]Emp. & \hspace{3mm}\begin{tabular}{@{}c@{}} $-21.22$dB\end{tabular}  & \hspace{3mm}\begin{tabular}{@{}c@{}}  $-20.37$dB\end{tabular}  & \hspace{3mm}\begin{tabular}{@{}c@{}}  $-19.57$dB\end{tabular}& \hspace{3mm}\begin{tabular}{@{}c@{}}  $-19.28$dB\end{tabular} \\ [0.0ex]
$|\Scal_{0:T}|$ & \hspace{3mm}\begin{tabular}{@{}c@{}} $277$\end{tabular}  & \hspace{3mm}\begin{tabular}{@{}c@{}}  $61$\end{tabular}  & \hspace{3mm}\begin{tabular}{@{}c@{}}  $37$\end{tabular}& \hspace{3mm}\begin{tabular}{@{}c@{}}  $32$\end{tabular}\\ 
 \bottomrule
\end{tabular}
\end{center}
\end{table}

First, we analyze the effect of the sampling set $\Scal_{0:T}$ when the graph process is perfectly $\Fcal-$bandlimited. In this regard, we considered $|\Fcal| = N = 32$ (i.e., the entire bandwidth) and corrupted the measurements with a zero-mean Gaussian noise with $\sigma_v^2 = 10^{-1}$, which corresponds to an average signal-to-noise ratio (SNR) of $19.3$dB computed as
\begin{equation}
\overline{\textnormal{SNR}} = 10\textnormal{log}_{10}\left[\frac{\sum_{\tau = 1}^{R}\|\r_\tau\|_2^2}{NR\sigma_v^2}\right].
\end{equation}
Here, $\r_\tau$ stands for the $\tau$th recording. The $|\Scal_{0:T}|$ samples are chosen by solving the opposite of problem \eqref{opt.mse_sel} as the ones that minimize the MSE \eqref{mse_theo1} in a sparse sense fashion. As a performance evaluation criterion, we use the normalized MSE (NMSE) between the estimated (observed) $\tau$th recording ${\r}_\tau^o$ and the true one $\r_\tau$, defined as
\begin{equation}
\textnormal{NMSE} = \frac{\sum_{\tau = 1}^{R}\|\r_\tau^o - \r_\tau	\|^2}{\sum_{\tau = 1}^{R}\|\r_\tau\|^2}.
\end{equation}

Fig.~\ref{fig.detObs1} shows the obtained NMSE as a function of $|\Scal_{0:T}|$. It can be seen that even with $60$ samples (out of $320$) an NMSE of $-20$dB is achieved. On the contrary, the uniformly random sampling\footnote{To account for the randomness in this sampling strategy the NMSE is further averaged over $100$ iterations.} requires far more measurements to give a comparable performance. \emph{This finding suggests that the sparse observability approach can also be implemented for graph processes that have a contribution on the entire bandwidth.}

{To provide more insights, Table~\ref{table:coeff} shows the theoretical and empirical NMSE as a function of the target value $\gamma$ in \eqref{opt.mse_sel}. In addition, we show also the cardinality of $\Scal_{0:T}$. We observe that a stricter NMSE requirement in \eqref{opt.mse_sel} leads to a higher $|\Scal_{0:T}|$ and, vice-versa, a bigger $\gamma$ leads to a sparser $\Scal_{0:T}$.}

\begin{figure}[t]
  \centering
    \includegraphics[width=0.5\textwidth, trim={.1cm 0 .25cm .33cm},clip]{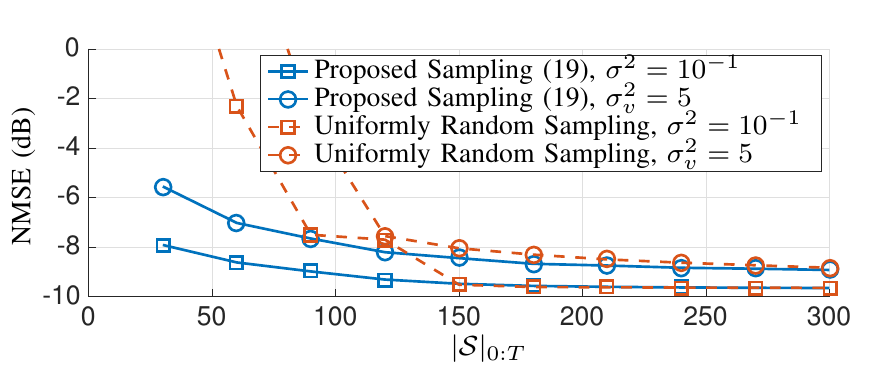}
    \vskip -.45cm
      \caption{NMSE versus the number of selected graph-time locations for the sampling algorithm \eqref{opt.mse_sel} and a uniformly random sampling. The state spectral evolution is considered localized on $\Fcal$ with $|\Fcal| = 8$.}
          \label{fig.detO_1}
\end{figure}

\begin{figure}[t]
  \centering
    \includegraphics[width=0.5\textwidth, trim={.1cm 0 .25cm .33cm},clip]{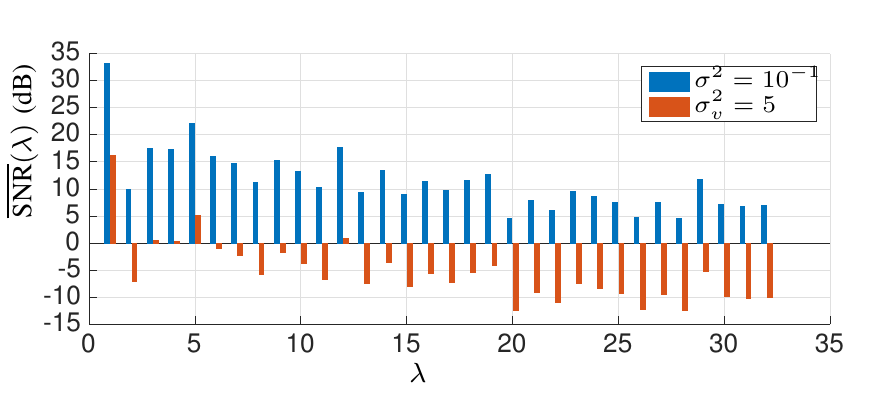}
    \vskip -.45cm
      \caption{Average SNR per graph frequency for the two different noise powers computed as $\overline{\textnormal{SNR}}(\lambda_n) = \sum_{\tau = 1}^{744}\hat{r}_n^2/T\sigma_v^2$. In the high noise regime, we observe that most frequencies experience a negative SNR.}
          \label{fig.avSNR}
\end{figure}

\begin{figure}[t]
  \centering
    \includegraphics[width=0.5\textwidth,trim={.2cm 0 .25cm .33cm},clip]{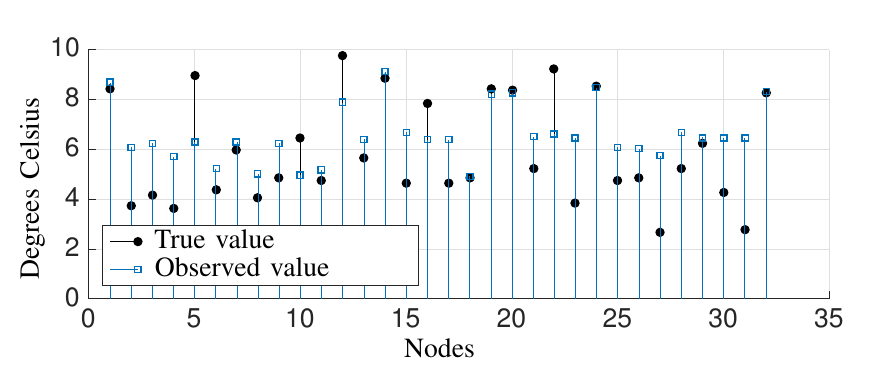}
        \vskip -.45cm
      \caption{True temperature values and the observed ones for a random recording; $|\Fcal| = 8$, $|\Scal_{0:T}| = 60$, and $\sigma_v^2 = 10^{-3}$. Further improvement can be obtained by increasing $|\Fcal|$.}
          \label{fig.detO_2}
\end{figure}

{In the second scenario, we restrict the process bandwidth to the first $|\Fcal| = 8$ graph frequencies and analyze two different noise variances $\sigma_v^2 = \{10^{-1}, 5\}$ ($\overline{\textnormal{SNR}} = \{19.3\textnormal{dB}, 2.3\textnormal{dB}\}$). The sampling set $\Scal_{0:T}$ is built as in the previous scenario and is again compared with the uniformly random sampling.

Fig.~\ref{fig.detO_1} depicts the average NMSE as a function of $|\Scal_{0:T}|$,} where the proposed selection strategy outperforms again the uniformly random sampling. We further observe that the NMSE has a lower floor much higher than for the full bandwidth case and its value does not reduce even by increasing $|\Scal_{0:T}|$. {We attribute this limitation to the restricted bandwidth, since the out-of-band signal contribution seems playing a role in improving further the performance.} In fact, w.r.t. Fig.~\ref{fig.avSNR}, we observe that in low noise regimes it is beneficial to consider a larger bandwidth since the average SNR per frequency is high. On the contrary, this might not be the case for $\sigma_v^2 = 5$, since the $\overline{\textnormal{SNR}}(\lambda_n)$ is negative for high graph frequencies. In the sequel, we show that indeed the $\overline{\textnormal{SNR}}(\lambda_n)$ plays a crucial role in the observability performance. Fig.~\ref{fig.detO_2} concludes this scenario by plotting the true signal and the corresponding observed signal with $|\Scal_{0:T}| = 60$ and $\sigma_v^2 = 10^{-1}$ for a random pick.

\begin{figure}[!]
  \centering
    \includegraphics[width=0.5\textwidth, trim={.1cm 0 .25cm .33cm},clip]{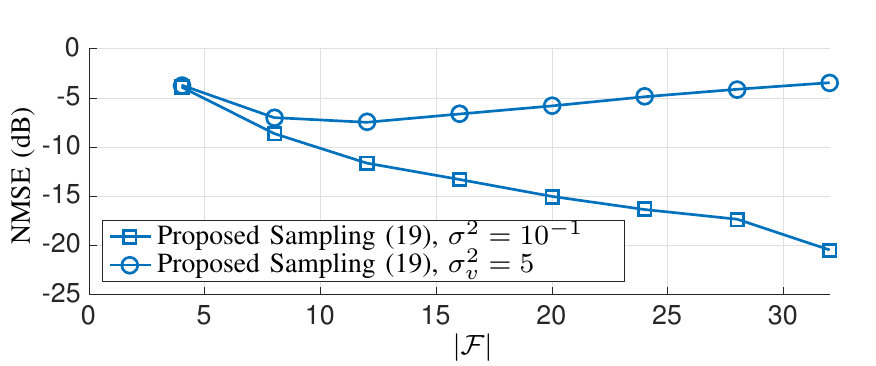}
        \vskip -.45cm
      \caption{NMSE versus the signal bandwidth $|\Fcal|$ for different noise powers. The sampling set has cardinality $|\Scal_{0:T}| = 100$ chosen by minimizing the MSE~\eqref{mse_theo1}. Observe that a larger bandwidth is not favorable when the measurement noise has high power.}
          \label{fig.detO_3}
\end{figure}

In this third scenario, we analyze the effects of the signal bandwidth on the observability performance. We fix $|\Scal_{0:T}| = 60$ samples (i.e., almost twice the full bandwidth) and compute the NMSE for different values of $|\Fcal|$ and $\sigma_v^2$. These results are shown in Fig.~\ref{fig.detO_3}.

We observe an increasing trend of the NMSE in high noise regimes (i.e., $\sigma_v^2 = 5$). This suggests that the meaningful information is concentrated in the first few frequencies and, therefore, the graph process is bandlimited. As highlighted in Fig.~\ref{fig.avSNR}, by increasing $|\Fcal|$ we only add more noise resulting in a performance degradation. \emph{This result suggests that in the presence of noise the process bandwidth should not be determined solely by the signal energy, but by the signal-to-noise ratio (SNR).} Indeed, a larger bandwidth (although the signal has energy content) degrades the overall $\overline{\textnormal{SNR}}$. This finding is further reinforced in the low noise regime, where a larger bandwidth is preferred to exploit the SNR on the high frequencies for better observing the graph process.

\begin{figure*}[!t]
\minipage{0.32\textwidth}
  \includegraphics[width=\linewidth,trim={2cm 0 1.35cm 0},clip]{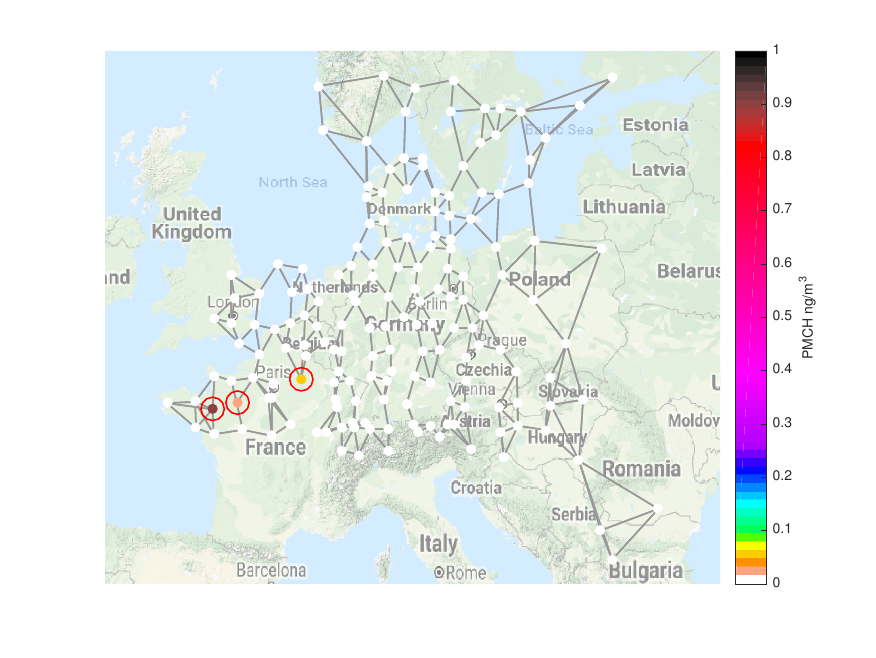}
\label{fig:awesome_image1}
\endminipage\hfill
\minipage{0.32\textwidth}
  \includegraphics[width=\linewidth,trim={2cm 0 1.35cm 0},clip]{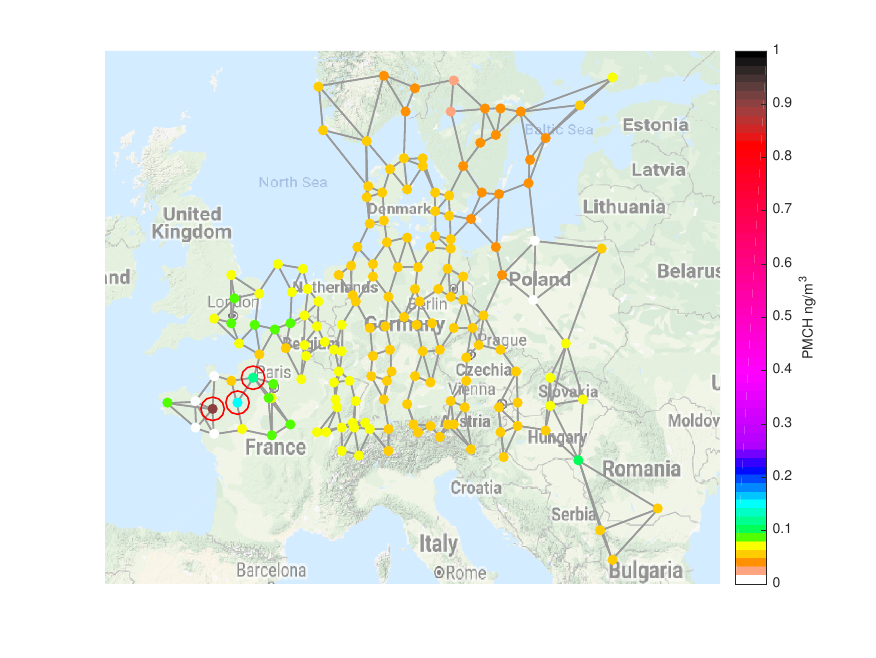}
\label{fig:awesome_image2}
\endminipage\hfill
\minipage{0.32\textwidth}%
  \includegraphics[width=\linewidth,trim={2cm 0 1.35cm 0},clip]{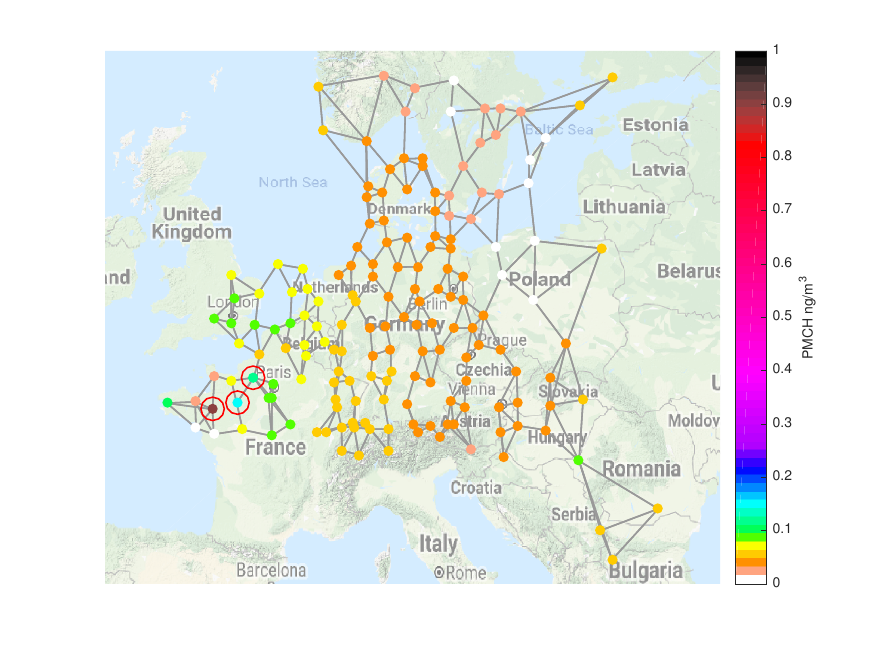}
  \label{fig:awesome_image2}
\endminipage
\vskip -1cm
\caption{Tracer (PMCH) concentration in the 168 stations. {The top three nodes with the highest PMCH concentration are circled in red. (Left) Ground truth concentration at $t = 0$. (Center) Observed tracer concentration by the intantaneous diffusion model with $w = 3.5$ on the $3$NN graph with all nodes collecting samples. (Right) Mean observed tracer concentration following the observability with random sampling and overall sampling rate $\mathbf{1}^{\transp}_N\bar{\c} = 60$ (out of $168$) by solving the opposite of problem \eqref{eq.opt_prob_obs}. The NMSE between the observed signal with random sampling (right) and the reconstructed ground truth (center) is $-16.2$ dB with a variance of $-48.1$dB around this value.}}\vspone\vspone\vspone\vspone\vspone
 \label{fig:Etex_maps}
\end{figure*}

\textbf{Obs. with random sampling.} The ETEX experiment \cite{nodop1998field} contains measurements of an identifiable perfluorocarbon concentration, released near Rennes, France, and then diffused over Europe.
Thirty concentration measurements were collected over a period of 72 hours at $N = 168$ ground-level stations. These stations will serve as the nodes of a $k$NN graph and the $30$ collected measurements in time will be the graph process. For several reasons, the measurements are not always available and in these cases, the tracer concentration is set to zero.
%

We considered model \eqref{eq.inst_diff} to capture the signal evolution over time. The set $\Fcal$ contains the frequency indices where $30\%$ of the process energy is concentrated ($|\Fcal| = 6$). Since diffused graph signals are often bandlimited, our intuition is that also in this experiment most of the frequencies will not have useful information. With this setup, we found heuristically that $k = 3$ and $w = 3.5$ lead to the smallest observability error by using all ($168\times72$) recordings. We then use this result to test the observability with random sampling. {To account for the measurement noise, the signal is corrupted with a zero-mean Gaussian noise of variance $\sigma_v^2 = 10^{-4}$ ($\overline{\textnormal{SNR}} = 29.2$dB). The obtained results are averaged over 2000 iterations.}

\begin{figure}[t]
  \centering
    \includegraphics[width=0.5\textwidth, trim={.1cm 0 .25cm .33cm},clip]{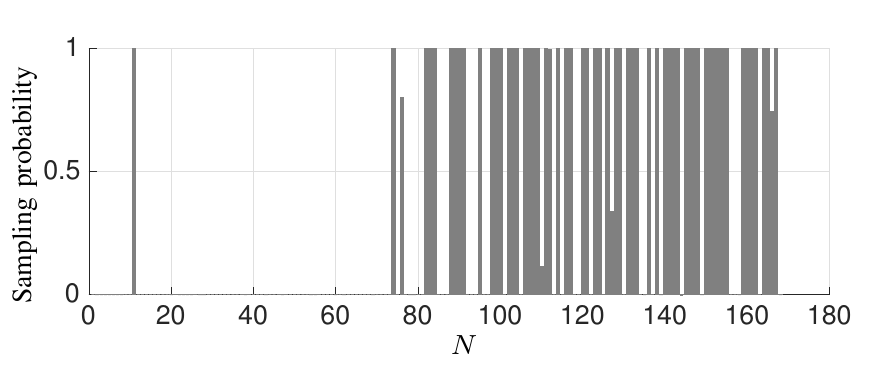}
    \vskip -.45cm
      \caption{Optimal sampling probabilities over the nodes obtained from solving the opposite of problem~\eqref{eq.opt_prob_obs}. We observe that several nodes are sampled with probability one and that the overall solution is highly sparse.}
          \label{fig.sampProb}
\end{figure}

In Fig.~\ref{fig:Etex_maps} (left), we plot the original signal at $t = 0$, wherein red circles highlight the top three nodes with the highest concentration. Then, in Fig.~\ref{fig:Etex_maps} (center) we plot the initial signal reconstructed by using model \eqref{eq.inst_diff} when all nodes collect data. {For this instance, the fitted graph and the used model are capable to identify the region (specifically the top two highest concentrations) where the tracer was released, but at the same time a tracer concentration around $0.04$ng/m$^3$ is also observed over all nodes\footnote{{We attribute this concentration leakage to the missing values that are set to zero and to the absence of wind information on the specific days.}}. However, for this work, we will use} this fully observed signal (Fig.~\ref{fig:Etex_maps} (center)) as a benchmark since it is the best that we can reconstruct with the fitted model.
%
%
%
{In Fig.~\ref{fig:Etex_maps} (right), we show the average observed signal with a sampling rate of $60$. The sampling probabilities are obtained by solving the opposite of problem \eqref{eq.opt_prob_obs} and are illustrated in Fig.~\ref{fig.sampProb}. We achieved an average NMSE between the observed signal and the reconstructed ground truth (Fig.~\ref{fig:Etex_maps} (center)) of $-16.2$dB with a variance around this performance of $-48.1$dB.}

\emph{{The above results lead to the following conclusions: $i)$ the deviation of a particular realization from the averaged observed signal is in general negligible, yielding a good practical result; and $ii)$ similarly to the approaches that use the CRLB to perform sparse sampling, the lower bound \eqref{eq.propMSE_eq2} is a suitable cost function to design a sparse sampler.}}

{Finally, in Table~\ref{table:RndObs} we address the impact of $\gamma$ in \eqref{eq.opt_prob_obs} on the lower bound \eqref{eq.propMSE_eq2}, the empirical NMSE, the overall sampling rate, and the value of $\alpha$ in \eqref{eq.prob}. We observe that despite the gap between the theoretical lower bound and the empirical NMSE, a looser requirement on \eqref{eq.propMSE_eq2} induces a lower sampling rate. Moreover, all the reported values of $\alpha$ lead to a probability \eqref{eq.prob} below machine precision. This demonstrates the use of \eqref{eq.propMSE_eq2} for sparse sampling design and that a reasonable NMSE is achieved even by collecting $1/3$ of the measurements.}\vspone\vspone\vspone\vspone

\setlength{\extrarowheight}{2.0pt}
\setlength{\tabcolsep}{9pt}
\begin{table}[t]
\linespread{1.5}
\centering
\caption{ Theoretical, empirical NMSE, overall sampling rate, and $\alpha$ in \eqref{eq.prob} for different values of $\gamma$ ($\times 10^{-4}$) in \eqref{opt.mse_sel}.
}\vskip-3mm
\label{table:RndObs}
\begin{center}
\vspace{-3mm}
 \begin{tabular}{@{}c @{}c @{}c @{}c @{}c@{}} \toprule
\textit{\ \ } & $\gamma = 3.12$ & $\gamma = 3.15$& $\gamma = 3.18$& $\gamma = 3.21$ \\  
 \midrule 
Theo. Lower Bound \eqref{eq.propMSE_eq2} & \hspace{3mm}\begin{tabular}{@{}c@{}} $-36.47$dB\end{tabular}  & \hspace{3mm}\begin{tabular}{@{}c@{}}  $-36.42$dB\end{tabular}  & \hspace{3mm}\begin{tabular}{@{}c@{}}  $-36.38$dB\end{tabular}& \hspace{3mm}\begin{tabular}{@{}c@{}}  $-36.34$dB\end{tabular} \\ [0.0ex]
\rowcolor{black!7}[0pt][0pt]Emp. & \hspace{3mm}\begin{tabular}{@{}c@{}} $-24.52$dB\end{tabular}  & \hspace{3mm}\begin{tabular}{@{}c@{}}  $-18.77$dB\end{tabular}  & \hspace{3mm}\begin{tabular}{@{}c@{}}  $-16.48$dB\end{tabular}& \hspace{3mm}\begin{tabular}{@{}c@{}}  $-16.13$dB\end{tabular} \\ [0.0ex]
$\mathbf{1}^{\transp}_N\bar{\c} $ & \hspace{3mm}\begin{tabular}{@{}c@{}} $130.4$\end{tabular}  & \hspace{3mm}\begin{tabular}{@{}c@{}}  $83.9$\end{tabular}  & \hspace{3mm}\begin{tabular}{@{}c@{}}  $62.7$\end{tabular}& \hspace{3mm}\begin{tabular}{@{}c@{}}  $50.2$\end{tabular}\\ [0.0ex]
$\alpha $ & \hspace{3mm}\begin{tabular}{@{}c@{}} $3910$\end{tabular}  & \hspace{3mm}\begin{tabular}{@{}c@{}}  $2515 $\end{tabular}  & \hspace{3mm}\begin{tabular}{@{}c@{}}  $1881$\end{tabular}& \hspace{3mm}\begin{tabular}{@{}c@{}}  $1505$\end{tabular}\\ 
 \bottomrule
\end{tabular}
\end{center}
\vsp\vsp\vsp
\end{table}

\subsection{Tracking graph processes}
\label{subsec.num_obs}

We now analyze the tracking performance of the KF approaches in Section~\ref{sec.kalman}. {We first consider KF for time-varying models and then focus on steady-state KF. The results are averaged over 500 different realizations.}

{\textbf{KF for time-varying models.}} We consider tracking instantaneous graph signal diffusion on the Molene data set. {The graph is a $3$NN, $\Fcal$ consists of the first $16$ graph frequencies, and $w = 1$ in equation \eqref{eq.inst_diff}.} \rev{The state $\x_{0}$ is initialized as zero and $\u_t$ for $t \in \{1, 101, \ldots, 401\}$ consists of five temperature recordings from the data set with $\B_t = \I_N$. In a nutshell, the state evolution considers the temperature diffusion for 100 iterations and then a new input is introduced.} {We consider a zero-mean model and measurement noises with respective covariance matrixes $\bSigma_w = 10^{-4}\I_N$ and $\bSigma_v = 10^{-1}\I_N$. We initialize the Kalman filter with $\tx_0^+ = \1_{|\Fcal|}$ and $\P_0^+ = \bSigma_{\tilde{w}}$.} We compare the sparse sensing approach (i.e., the opposite problem of \eqref{eq.KF_nodeS} that selects $|\Scal|$ nodes with minimum MSE) and uniformly random sampling whose performance is averaged over $500$ additional realizations.

Fig.~\ref{fig.KF_track} illustrates the tracking performance as a function of the iteration index for different values of $|\Scal_t|$. We observe that an increment of $|\Scal_t|$ leads to a smaller NMSE, {especially in the first iterations}. However, compared to the case of full bandwidth and $|\Scal|_t = 32$, these results show that $50\%$ of the samples can be saved by the proposed approach with a little tradeoff on the NMSE. Further, as in \cite{chepuri2016sparse}, uniformly random sampling can be an option for tracking the process for large $t$. {We additionally remark that \eqref{eq.KF_nodeS} may not always give a sparse solution for higher $t$ and since it is an SDP relaxation, it might often lead to solutions that are far from the possible minimum MSE. Finally, note that the spikes in the estimated NMSE are related to the presence of the input signal and are common for both sampling approaches. 

\begin{figure}[t]
  \centering
    \includegraphics[width=0.5\textwidth]{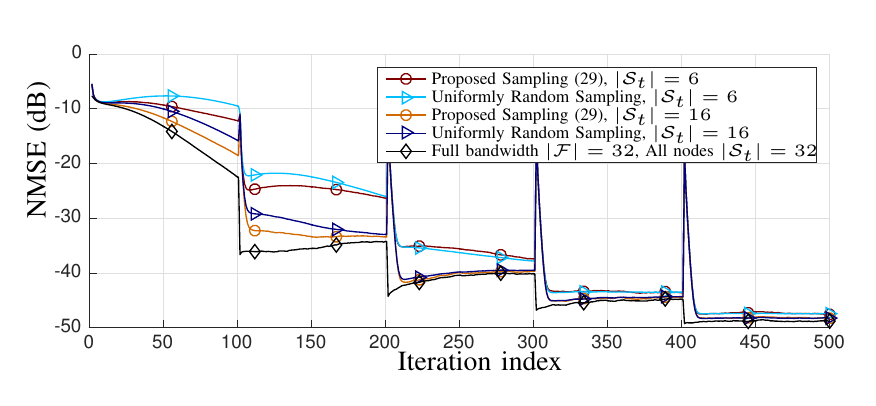}\vskip -.45cm\vsp\vsp\vsp\vsp\vsp
      \caption{Tracking performance of KF. Estimated NMSE versus iteration index for different numbers of sampled nodes. The results are analyzed for the sampling approach \eqref{eq.KF_nodeS}, uniformly random sampling, and when all the nodes are sampled.}\vspone
         \label{fig.KF_track}
\end{figure}

\begin{figure}[t]
  \centering
    \includegraphics[width=0.5\textwidth]{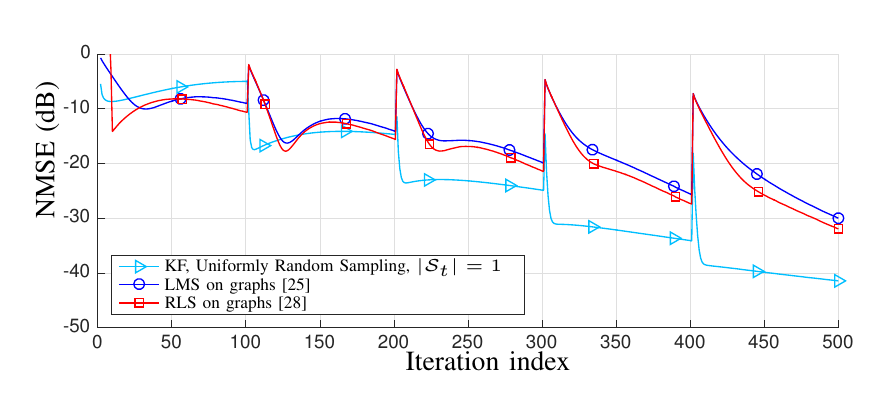}\vskip -.45cm\vsp\vsp\vsp\vsp\vsp
      \caption{Estimated NMSE versus iteration index for KF, LMS ($\mu = 0.125$) \cite{PDL2016_TSIPN} and RLS ($\beta = 0.95$) \cite{di2017adaptive}. For KF, one node is sampled for each iteration, while LMS and RLS have an average sampling rate of $16.08$ (greater than $|\Fcal| = 16$) with five nodes sampling with probability one.}
         \label{fig.rls_kf_lms_track}
\end{figure}


Next, we compare the tracking performance of KF with that of LMS \cite{PDL2016_TSIPN} and RLS \cite{di2017adaptive} on graphs. {For the KF approach $\Scal_t$ consists of one node, sampled at random for each $t$. The RLS sampling probabilities are found with $\beta_{\text{RLS}} = 0.95$ and $\gamma_{\text{RLS}} = 7\times10^{-2}$ following the optimal design of \cite{di2017adaptive}. The latter results in an average sampling rate of $16.08$ (greater than $|\Fcal| = 16$) for each $t$, with five nodes sampled with probability one. With the same sampling probabilities, the LMS step size is $\mu_{\text{LMS}} = 0.0875$ such that it meets the RLS steady-state MSE. Both algorithms are initialized as the KF.}

{The results of Fig.~\ref{fig.rls_kf_lms_track} show that the KF suffers only in the first iterations, but as the system evolution is learned better it outperforms both the LMS and RLS and, as a consequence, other state-of-the-art tracking algorithms \cite{PDL2016_TSIPN,didistributed,di2017adaptive} with which LMS and RLS compare.} This result highlights the potential of the proposed approach to optimally track the signal by sampling only one node per time instant, while exploiting its dynamics. 

{\textbf{Steady-state KF.} We now track a heat diffusion process evolving on a binary weighted two-dimensional rectangular grid of $N = 75$ nodes ($5 \times 15$) by making use of the steady-state KF approach. Here, we aim at providing insights into how GSP can be exploited in temperature monitoring systems. The initial signal $\x_0$ is set to one at the five nodes of the leftmost column of the grid and zero elsewhere. This signal is diffused following the heat propagating model \eqref{eq.inst_diff} with $w = 10$ for $T = 500$ instances. $\Fcal$ consists of the frequency indices where $99\%$ of the energy of $\x_0$ is concentrated, resulting in $|\Fcal| = 18$ active frequencies (not necessarily adjacent). The model and measurement noises have  covariance matrices $\bSigma_w = 10^{-4}\I_N$ and $\bSigma_v = 10^{-1}\I_N$, respectively.
}

\begin{figure}[t]
  \centering
    \includegraphics[width=0.5\textwidth]{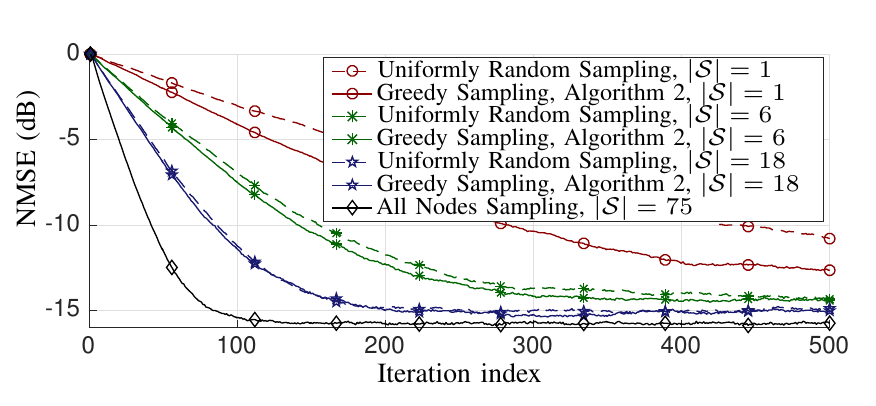}\vskip -.45cm\vsp\vsp\vsp\vsp\vsp
      \caption{Estimated NMSE versus iteration index for steady-state KF with different numbers of sampled nodes. The results are analyzed for the sampling approach in Algorithm \ref{alg_greedyssKF}, uniformly random sampling, and when all the nodes are sampled.}\vspone
         \label{fig.SS_KF_Track}
\end{figure}

\begin{figure}[t]\vsp\vsp\vsp
  \centering
    \includegraphics[width=0.5\textwidth]{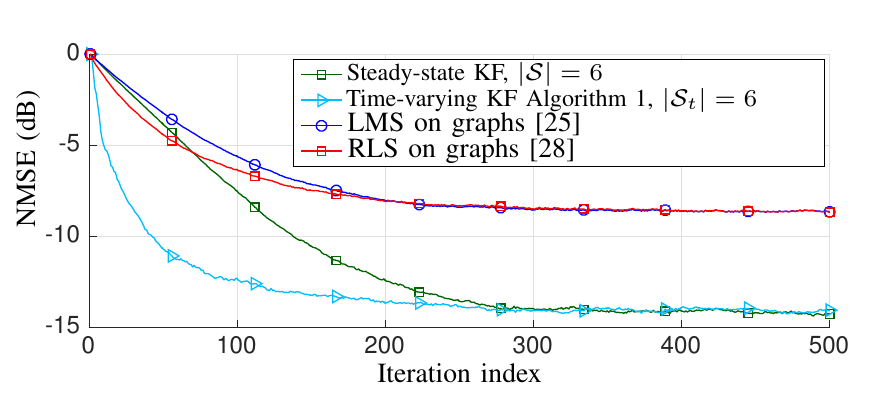}\vskip -.45cm\vsp\vsp\vsp\vsp\vsp
      \caption{Estimated NMSE versus iteration index for steady-state KF ($|\Scal| = 6$ chosen with Algorithm~\ref{alg_greedyssKF}), time-varying KF ($|\Scal_t| = |\Scal| = 6$ chosen randomly for each iteration), as well as RLS ($\beta_{\text{RLS}} = 0.99$) and LMS ($\mu_{\text{LMS}} = 0.041$) with maximum sampling rate $|\Scal_t| = |\Fcal| = 18$.}
         \label{fig.SS_KF_Comp}
\end{figure}

{Fig.~\ref{fig.SS_KF_Track} shows the NMSE as a function of the diffusion time for different cardinalities of the sampling set. We observe that a larger $|\Scal|$ improves the steady-state NMSE and the convergence rate. Additionally, the sampling strategy in Algorithm~\ref{alg_greedyssKF} is beneficial for low values of $|\Scal|$, while the uniformly random sampling can only be adopted for larger $|\Scal|$.}

Finally, we compare the tracking performance of the steady-state KF with LMS and RLS on graphs. The LMS and RLS parameters are chosen as before yielding $\mu_{\text{LMS}} = 0.041$ (LMS), $\beta_{\text{RLS}} = 0.99$ (RLS), and an overall sampling rate of 18 samples per iteration (i.e., the same as $|\Fcal|$ to guarantee asymptotic MSE reconstruction) \cite{di2017adaptive}. Additionally, KF with random time-varying sampling from Algorithm~\ref{alg_kf} is considered as a benchmark. The instantaneous sampling set $\Scal_t$ is chosen uniformly at random for each $t$ with $|\Scal_t| = |\Scal| = 6$.


{The results in Fig.~\ref{fig.SS_KF_Comp} show that the KF strategies outperform the adaptive algorithms in both steady-state performance and convergence speed. We also remark that the steady-state performance of both KF approaches is identical with a sampling rate that is three times lower than the other alternatives. The additional improvement in convergence speed of the time-varying KF comes at the expense of complexity, i.e., updating the Kalman gain matrix and the \emph{a priori} and \emph{a posteriori} error covariance matrices in each iteration.}\vspone\vspone\vspone
%
%
%
%
%
%
%
%
%

\section{Conclusion}
\label{sec.concl}

This work proposed strategies to observe and track bandlimited graph processes. 
We first merged observability concepts with graph signal processing for which we derived necessary and sufficient conditions to observe a bandlimited graph process from a subset of nodes. Further, we introduced the idea of observability with random sampling, where the nodes are sampled with a given probability. Also for the latter case, we derived conditions for observing the graph process and proposed a novel way to design the sampling probabilities in a sparse sensing fashion. 
Next, we proposed Kalman filtering for tracking bandlimited graph processes given some knowledge on their system evolution. We derived conditions for the minimum number of nodes that should be sampled to fully exploit the Kalman filter. Finally, we provided sampling strategies that ensure a target tracking performance for both finite and steady-state performance.
Different numerical tests corroborated our findings and show that exploiting the bandlimited prior reduces drastically the number of sampled nodes.

\rev{Future research should be focused on two main directions. First, to analyze the effects of graph learning algorithms on the sampling strategy; and second, to extend the framework to epidemic diffusion.}
\vspone\vspone\vspone\vspone

\section{Appendix}

\vskip 5mm\vspone\vspone\vspone
\emph{A) Proof of Proposition~\ref{eq.prop_det_observability}}
\vskip 2mm\vsp
By applying the rank inequality
\begin{equation}
\text{rank}(\A\B) \le \text{min}\{	\text{rank}(\A), \text{rank}(\B)	\}
\end{equation}
to $\O_{{0:T}} =  \C_{\Scal_{0:T}}(\I_{T+1}\otimes\U_\Fcal)\tA_{0:T}$ in \eqref{eq.obsmat}, we have that $\O_{0:T}$ can be full column rank $|\Fcal|$ only if
\begin{equation}
\label{eq.obs_condition_Fin}
\text{rank}\left(\C_{\Scal_{0:T}}	\right) \ge |\Fcal|,
\end{equation}
which from the structure of $\C_{\Scal_{0:T}}$ is always true when the claimed conditions are satisfied. \qed

\vskip 5mm\vspone\vspone\vspone
\emph{B) Proof of Theorem~\ref{theo_obs}}
\vskip 2mm

By substituting $\C_{\Scal_{0:T}} = \I_{N(T+1)} - \C_{\Scal_{0:T}}^c$ into the rank argument of \eqref{eq.ranktF1} we can write the vector form expression
\begin{align}
\label{eq.com_set}
\begin{split}
&\tA_{0:T}^{\herm}(\I_{T+1}\otimes\U_\Fcal^{\herm})\C_{\Scal_{0:T}}(\I_{T+1}\otimes\U_\Fcal)\tA_{0:T} = \tA_{0:T}^{\herm}\tA_{0:T} \\
&\quad- \tA_{0:T}^{\herm}(\I_{(T+1)}\otimes\U_\Fcal^{\herm})\C_{\Scal_{0:T}^c}(\I_{T+1}\otimes\U_\Fcal)\tA_{0:T},
\end{split}
\end{align}
which is invertible if
\begin{equation}
\label{eq.proof_th1_dymmy0}
\|\tA_{0:T}^{\herm}\!(\I_{T\!+\!1}\otimes\U_\Fcal^{\herm})\C_{\Scal_{0:T}^c}(\I_{T\!+\!1}\otimes\U_\Fcal)\tA_{0:T}\| \!<\! \lambda_\text{min}(\tA_{0:T}^{\transp}\tA_{0:T}),
\end{equation}
where $\lambda_{\text{min}}(\A)$ is the minimum eigenvalue of $\A$. Here, we are exploiting that both matrices on the right-hand side of \eqref{eq.com_set} are positive semidefinite. Then, from the Cauchy-Schwarz inequality we have
\begin{align*}
\begin{split}
&\|\tA_{0:T}^{\herm}(\I_{T\!+\!1}\otimes\U_\Fcal^{\herm})\C_{\Scal_{0:T}^c}\!(\I_{T\!+\!1}\otimes\U_\Fcal)\tA_{0:T}\| \!\le\!\\
&~ \|(\I_{T\!+\!1}\!\otimes\!\U_\Fcal^{\herm})\| \|\C_{\Scal_{0:T}^c}(\I_{T\!+\!1}\!\otimes\!\U_\Fcal)\| \|\tA_{0:T}\|^2 \!<\! \lambda_\text{min}(\tA_{0:T}^{\herm}\tA_{0:T}),
\end{split}
\end{align*}
which then leads to ($ \|(\I_{T\!+\!1}\!\otimes\!\U_\Fcal^{\transp})\| = 1$)
\begin{equation}
\label{eq.proof_th1_dymmy1}
\|\C_{\Scal_{0:T}^c}(\I_{T\!+\!1}\!\otimes\!\U_\Fcal)\| \!<\! \frac{\lambda_\text{min}(\tA_{0:T}^{\transp}\tA_{0:T})}{ \|\tA_{0:T}\|^2} = \frac{s^2_{\text{min}}(\tA_{0:T})}{s^2_{\text{max}}(\tA_{0:T})}.
\end{equation}
The equality in \eqref{eq.proof_th1_dymmy1} derives from the definition of the spectral norm and the relation between the singular and the eigenvalues of a matrix. To prove that \eqref{eq.theo2} is a neccessary and sufficient condition we follow similar arguments as in \cite{tsitsvero2016signals,PDL2016_TSIPN}. From \eqref{eq.proof_th1_dymmy0}, $\O_{0:T}$ is full rank if the sufficient condition \eqref{eq.theo2} holds. Conversely, if $\|\C_{\Scal_{0:T}^c}(\I_{T\!+\!1}\!\otimes\!\U_\Fcal)\| = \lambda_\text{min}(\tA_{0:T}^{\herm}\tA_{0:T})/{ \|\tA_{0:T}\|^2}$ for $T = 0$ and thus $\tA_{0:0}= \I_N$ we have $\|\C_{\Scal_{0}^c}(\I_{1}\!\otimes\!\U_\Fcal)\| = 1$ which goes in contradiction with the conventional observability (recovery) of bandlimited graph signals (e.g., \eqref{eq.rec_x}). This proves that \eqref{eq.theo2} is also necessary. \qed

\vskip 5mm\vspone\vspone\vspone
\emph{C) Proof of Proposition~\ref{prop.obs_mean} }
\vskip 2mm
From the structure of $\C_{\Scal_{0:T}}$, we have
\begin{equation}
\label{eq.rnd_proof1}
\text{rank}(\C_{\Scal_{0:T}}) \le \text{rank}(\Exp[\C_{\Scal_{0:T}}]) = \text{rank}(\I_{T+1}\otimes\bC).
\end{equation}
A necessary condition then for $\text{rank}(\C_{\Scal_{0:T}})$ to be $|\Fcal|$ is that
\begin{equation}
\label{eq.rnd_proof2}
\text{rank}(\I_{T+1}\otimes\bC) \ge |\Fcal|.
\end{equation}
From $\text{rank}(\A\otimes\B) = \text{rank}(\A)\text{rank}(\B)$, \eqref{eq.rnd_proof2} writes as
\begin{equation}
\label{eq.rnd_proof3}
\text{rank}(\bC) \ge {|\Fcal|}/(T+1).
\end{equation}
Then, since $\bC$ is diagonal means that at least $\lceil|\Fcal|/(t+1)\rceil$ nodes must be sampled with a probability different from zero. The latter concludes the proof. \qed

\vskip 5mm\vspone\vspone\vspone
\rev{\emph{D) Proof of Corollary~\ref{eq.corr_prob} }
\vskip 2mm}

\rev{Denote by $\c_t = \text{diag}(\C_{\Scal_t})$ the random sampling vector with expectation $\bar{\c}$ for $t \in \{0,\ldots,T\}$. Let also $d = |{\Scal_{0:T}}| = \sum_{t = 0}^{T}\sum_{n = 1}^Nc_{t,n}$ be an auxiliary variable that characterises the cardinality of the instantaneous sampling set ${\Scal_{0:T}}$. Then, $d$ is a Poisson random variable being it the sum of $N(T+1)$ independent Bernoulli random variables. The claim \eqref{eq.prob} follows by simple statistical properties.}

\vskip 5mm\vspone\vspone\vspone
{\emph{E) Proof of Proposition~\ref{prop_mseRndObs}}}

{By rewriting the MSE as
\begin{equation}
\MSE = \Exp_{\C}\left\{\Exp_{\v}\left[\tr\left[	(\tx_0^o - \tx_0)(\tx_0^o - \tx_0)^{\herm}	\right]	\right]\right\},
\end{equation}
from \eqref{mse_theo1} we have that
\begin{equation}\label{eq_appMSERnd1}
\MSE\!=\!\sigma_v^2\Exp_{\C}\!\left\{\!\tr\!\left[\!\left(\!\tA_{0:T}^{\herm}(\I_{T+1}\! \otimes\! \U_\Fcal)^{\herm}\C_{\Scal_{0:T}}(\I_{T+1}\! \otimes\! \U_\Fcal)\tA_{0:T}\!\right)^{\!\!-1}\!\right]\!	\right\}.
\end{equation}
Then, since the function $\varphi : \X \to \tr[\X^{-1}]$ is convex, we apply the Jensen inequality $\varphi(\Exp[\X]) \le \Exp[\varphi(\X)]$ to lower bound \eqref{eq_appMSERnd1} as in \eqref{eq.propMSE_eq2}.
\qed
}

\bibliographystyle{IEEEtran}
\bibliography{bibliography}

\flushend

\end{document}